\documentclass[lettersize,twoside,journal]{IEEEtran}
\usepackage{graphicx}
\usepackage{amssymb}
\usepackage{amsmath,cases}
\usepackage{cite}
\usepackage{xcolor}
\usepackage{subfigure}
\usepackage{booktabs}
\usepackage{balance}
\usepackage{acro}
\usepackage{stfloats}
\usepackage{multirow}
\usepackage{mathtools}
\usepackage[mathscr]{euscript}

\usepackage{xcolor}
\usepackage{braket}
\usepackage{bbm}
\usepackage{pifont} 
\usepackage{microtype}

\DeclareMathAlphabet{\mathpzc}{OT1}{pzc}{m}{it}

\DeclareMathAlphabet{\mathitsf}{OML}{cmbr}{m}{it}
\DeclareMathAlphabet{\mathsf}{OT1}{cmbr}{m}{n}

\definecolor{green1}{RGB}{0,51,51}
\definecolor{green2}{RGB}{51,102,102}
\definecolor{green3}{RGB}{125,180,180}
\definecolor{green4}{RGB}{180,218,218}

\definecolor{green0}{RGB}{0,25,25}      
\definecolor{green5}{RGB}{220,238,238}  

\definecolor{red1}{RGB}{220,20,60}  
\definecolor{red2}{RGB}{219,53,47}  

\definecolor{darkgreen}{RGB}{0,128,0}

\newtheorem{definition}{Definition}
\newtheorem{theorem}{Theorem}
\newtheorem{lemma}{Lemma}

\newtheorem{property}{Property}

\AtBeginDocument{}

\newcommand{\size}[1]{\ensuremath{|#1|}}

\newcommand{\lra}[1]{\ensuremath{(#1)}}
\newcommand{\lrb}[1]{\ensuremath{[#1]}}
\newcommand{\lrc}[1]{\ensuremath{\{#1\}}}
\newcommand{\lrd}[1]{\ensuremath{\langle #1\rangle}}

\newcommand{\lrA}[1]{\ensuremath{\left(#1\right)}}
\newcommand{\lrB}[1]{\ensuremath{\left[#1\right]}}
\newcommand{\lrC}[1]{\ensuremath{\left\{#1\right\}}}

\usepackage{makecell}

\allowdisplaybreaks

\usepackage{comment}

\usepackage[hidelinks]{hyperref}
\usepackage{algorithm}
\usepackage{algorithmic}

\usepackage{booktabs}
\usepackage{fixmath}
\usepackage{bm}
\usepackage{amsbsy}
\usepackage{color}
\usepackage{verbatim}
\usepackage{array}
\usepackage{tikz}
\usetikzlibrary{positioning}

\def\CC{\mathcal{C}}

\def\E{\mathcal{E}}
\def\F{\mathcal{F}}
\def\G{\mathcal{G}}

\def\O{\mathcal{O}}
\def\P{\mathcal{P}}

\def\S{\mathcal{S}}
\def\T{\mathcal{T}}

\def\V{\mathcal{V}}
\def\WW{\mathcal{W}}

\def\c{\hat{c}}

\DeclareAcronym{aas}
{
    short = a.a.s.,
    long = asymptotically almost surely
}
\DeclareAcronym{AP}{
short = AP,
long = a priori
}
\DeclareAcronym{ARP}{
short = ARP,
long = airports and railways problem
}
\DeclareAcronym{BTSP}{
short = BTSP,
alt = biased TSP,
long = biased traveling salesman problem
}
\DeclareAcronym{BTSPP}{
short = BTSPP,
long = biased traveling salesman path problem
}
\DeclareAcronym{CARP}{
short = CARP,
long = capacitated arc routing problem
}

\DeclareAcronym{CDCLRP}{
short = CD-CLRP,
alt = cross-docking CLRP,
long = cross-docking capacitated location routing problem
}

\DeclareAcronym{CLRP}{
short = CLRP,
long = capacitated location routing problem
}
\DeclareAcronym{CPP}{
short = CPP,
long = chinese postman problem
}

\DeclareAcronym{CVRP}{
short = CVRP,
alt = capacitated VRP,
long = capacitated vehicle routing problem
}

\DeclareAcronym{CuVRP}{
short = Cu-VRP,
alt = cumulative VRP,
long = cumulative vehicle routing problem
}

\DeclareAcronym{CuVRPSD}{
short = Cu-VRPSD,
alt = Cu-VRP with stochastic demands,
long = cumulative vehicle routing problem with stochastic demands
}

\DeclareAcronym{DaRP}{
short = DaRP,
long = dial-a-ride problem
}

\DeclareAcronym{DVRP}{
short = DVRP,
alt = distance constrained VRP,
long = distance constrained vehicle routing problem
}

\DeclareAcronym{EPTAS}{
short = EPTAS,
long = efficient PTAS
}

\DeclareAcronym{FLP}{
short = FLP,
long = facility location problem
}
\DeclareAcronym{FPT}{
short = FPT,
long = fixed-parameter tractable
}
\DeclareAcronym{FPTAS}{
short = FPTAS,
long = fully PTAS
}

\DeclareAcronym{GCLRP}{
short = G-CLRP,
alt = grouping CLRP,
long = grouping capacitated location routing problem
}

\DeclareAcronym{HetCVRP}{
short = HetCVRP,
alt = heterogeneous CVRP,
long = heterogeneous capacitated vehicle routing problem
}

\DeclareAcronym{ITP}{
short = ITP,
long = iterated tour partition
}

\DeclareAcronym{LP}{
short = LP,
long = linear programming
}

\DeclareAcronym{MO}{
short = MO,
long = merge operation
}
\DeclareAcronym{MDDaRP}{
short = MD-DaRP,
alt = multiple-depot DaRP,
long = multiple-depot dial-a-ride problem
}
\DeclareAcronym{MDSCP}{
short = MD-SCP,
alt = multiple-depot SCP,
long = multiple-depot stacker-crane problem
}
\DeclareAcronym{MDSDVRP}{
short = MD-SDVRP,
long = multiple-depot split delivery vehicle routing problem
}
\DeclareAcronym{MDTSP}{
short = MD-TSP,
alt = multiple-depot TSP,
long = multiple-depot traveling salesman problem
}
\DeclareAcronym{MDVRP}{
short = MD-VRP,
alt = multiple-depot VRP,
long = multiple-depot vehicle routing problem
}
\DeclareAcronym{MMCCP}{
short = MMCCP,
long = min-max cycle cover problem
}
\DeclareAcronym{MMTCP}{
short = MMTCP,
long = min-max tree cover problem
}
\DeclareAcronym{MST}{
short = MST,
long = minimum-weight spanning tree
}

\DeclareAcronym{MWSCP}{
short = MW$k$SCP,
long = minimum weight $k$-set cover problem
}

\DeclareAcronym{PCB}{
short = PCB,
long = printed circuit board
}
\DeclareAcronym{P2P}{
short = P2P,
long = point-to-point
}
\DeclareAcronym{PCCLRP}{
short = PC-CLRP,
alt = prize-collecting CLRP,
long = prize-collecting capacitated location routing problem
}

\DeclareAcronym{PTAS}{
short = PTAS,
long = polynomial-time approximation scheme
}

\DeclareAcronym{QPTAS}{
short = QPTAS,
long = quasi PTAS
}

\DeclareAcronym{RSAP}{
short = RSAP,
long = ride-sharing assignment problem
}
\DeclareAcronym{RPP}{
short = RPP,
long = rural postman problem
}

\DeclareAcronym{RS}{
short = RS,
long = restocking
}

\DeclareAcronym{RVRP}{
short = RVRP,
alt = regret-bounded VRP,
long = regret-bounded vehicle routing problem
}

\DeclareAcronym{SGPP}{
short = SGPP,
long = the subgroup path planning problem
}
\DeclareAcronym{SPP}{
short = SPP,
long = the subpath planning problem
}

\DeclareAcronym{SAT}{
short = SAT,
long = the satisfiability problem
}
\DeclareAcronym{SBP}{
short = SBP,
long = school bus problem
}

\DeclareAcronym{SCP}{
short = SCP,
long = stacker-crane problem
}
\DeclareAcronym{STP}{
short = STP,
long = Steiner tree problem
}

\DeclareAcronym{TRP}{
short = TRP,
long = traveling repairman problem
}
\DeclareAcronym{TSP}{
short = TSP,
long = traveling salesman problem
}
\DeclareAcronym{TSPP}{
short = TSPP,
long = traveling salesman path problem
}
\DeclareAcronym{TSPPK}{
short = $k$-TSPP,
long = $k$-person traveling salesman path problem
}
\DeclareAcronym{UITP}{
short = UITP,
alt = unsplittable ITP,
long = unsplittable iterated tour partition
}

\DeclareAcronym{VRP}{
short = VRP,
long = vehicle routing problem
}
\DeclareAcronym{VRPCC}{
short = VRPCC,
alt = VRP with compatibility constraints,
long = vehicle routing problem with compatibility constraints
}
\DeclareAcronym{EVRPRS}{
short = EVRPRS,
alt = Euclidean VRP with random setting,
long = Euclidean vehicle routing problem with random setting
}
\DeclareAcronym{VRPSD}{
short = VRPSD,
long = vehicle routing problem with stochastic demands
}

\DeclareAcronym{VRPTW}{
short = VRPTW,
alt = VRP with time windows,
long = vehicle routing problem with time windows
}

\DeclareAcronym{XP}{
short = XP,
long = slice-wise polynomial-time
}

\begin{document}

\title{
Constant-Factor Approximations for the Biased Traveling Salesman Problem and Its Path Variant
}
\author{
Jingyang~Zhao,
Jason William Setiawan,
Uman Khalid,
Chaoning~Zhang,~\IEEEmembership{Senior~Member,~IEEE},
Een-Kee~Hong,~\IEEEmembership{Senior~Member,~IEEE},
and
Hyundong~Shin,~\IEEEmembership{Fellow,~IEEE}
 \thanks{
J.~Zhao,
J.~Setiawan,
U.~Khalid,
E.-K.~Hong,
and H.~Shin (corresponding author)
 are with the Department of Electronics and Information Convergence Engineering,
 Kyung Hee University,
 1732 Deogyeong-daero, Giheung-gu,
 Yongin-si, Gyeonggi-do 17104, Republic of Korea
 (e-mail: hshin@khu.ac.kr).
 }
\thanks{
C.~Zhang is with
the School of Computer Science and Engineering, University of Electronic Science and Technology of China, Chengdu, China (e-mail: chaoningzhang@uestc.edu.cn).
}
 }
 \markboth{}{
 Zhao \textit{\MakeLowercase{et al.}}:
Constant-Factor Approximations for the Biased Traveling Salesman Problem and Its Path Variant
 }

\maketitle

\begin{abstract}
The metric \ac{TSP} is a fundamental problem in combinatorial optimization that asks for a minimum-cost Hamiltonian cycle in a complete undirected graph with $N$ vertices, where edge costs satisfy the triangle inequality. The \aca{BTSP} (\ac{BTSP}) is a natural generalization of the metric \ac{TSP}, in which the vertex set is partitioned into $k$ subsets: edge costs satisfy the triangle inequality within each subset, while edges crossing the subsets may violate the triangle inequality by a certain bias factor. This model captures scenarios in which distances are metric only locally but may be distorted across different regions or types of vertices. In this paper, we study approximation algorithms for the \ac{BTSP} and its path variant, the \acl{BTSPP} (\ac{BTSPP}), which asks for a minimum-cost Hamiltonian path. Previously, constant-factor approximation algorithms were known only for the case $k=2$. Whether the \ac{BTSP} and the \ac{BTSPP} admit constant-factor approximation algorithms for any fixed integer $k > 2$ has remained an open question.

We resolve this question affirmatively. We show that both the \ac{BTSP} and the \ac{BTSPP} admit $O\lra{k}$-approximation algorithms with running time $N^{O\lra{2^kk!}}$. Moreover, for the \ac{BTSPP} with $k=2$, we propose an improved $\lra{\alpha+2+\varepsilon}$-approximation algorithm, where $\alpha<3/2-10^{-34}$ is the best-known approximation ratio for the metric \ac{TSP}, improving upon the previous ratio of $4$.
\end{abstract}

\begin{IEEEkeywords}
The Traveling Salesman Problem, Approximation Algorithms, Biased Traveling Salesman Problem, Combinatorial Optimization.
\end{IEEEkeywords}





\section{Introduction}
\IEEEPARstart{R}{outing} and scheduling are fundamental optimization tasks in intelligent transportation systems, autonomous robotics, and large-scale cyber-physical networks~\cite{BMMD:24:IEEE_J_ITS,KOHTZ:15:IEEE_J_ITS,LXHZYI:23:IEEE_J_EVC,YHHW:19:IEEE_J_EVC,WCLY:16:IEEE_J_ITS,LIY:24:IEEE_J_ITS,ZKPZZS:26:IEEE_O_CSTO}. 
Efficient routing algorithms play a central role in applications ranging from logistics planning and mobile robot coordination to communication scheduling and manufacturing automation~\cite{ZVCZ:26:IEEE_J_ITS,LDNXW:24:IEEE_J_ITS,LGN:21:IEEE_J_ITS,ZZCSTZWD:2022:IEEE_J_ITS,LZC:23:IEEE_J_ITS,CLXLEC:24:IEEE_J_ITS,FHTGTTO:20:IEEE_J_ITS,ZQL:19:IEEE_J_ITS,JZWMCL:25:IEEE_J_ITS,NRSTD:25:IEEE_J_ITS}. 
Among the numerous routing models, the \acl{TSP} (\aca{TSP}) has long served as a fundamental optimization problem in combinatorial optimization and approximation algorithms~\cite{TV:2024:book,WS:2011:book,GJ:1979:book,MLDD:17:IEEE_J_ITS,KJKJJ:23:IEEE_J_ITS,AA:18:IEEE_J_ITS}. 
Given a complete graph $\G=\lrA{\V,\E}$ with non-negative edge costs, the goal is to compute a minimum-cost Hamiltonian cycle visiting every vertex exactly once~\cite{CLRS:2022:book,ABC:2006:book}.

The \ac{TSP} is NP-hard and cannot be approximated within any computable function unless $\mathscr{P}=\mathscr{NP}$~\cite{SG:76:JACM}. 
Nevertheless, when the edge-cost function satisfies the triangle inequality, the resulting metric \ac{TSP} admits polynomial-time approximation algorithms with provable performance guarantees. Research on the metric \ac{TSP} has progressed from classical algorithms, including the well-known $3/2$-approximation algorithm~\cite{C:2022:article,WS:2011:book}, to recent breakthroughs that achieve approximation ratios strictly less than $3/2$~\cite{KKG:21:stoc,KKG:23:ipco,LNJ:24:ICALP}. Throughout this paper, we use $\alpha< 3/2-10^{-34}$ to denote the best-known approximation ratio for the metric \ac{TSP}~\cite{LNJ:24:ICALP}.

\subsection{Background and Motivation}

Despite its broad applicability, the metric assumption does not accurately capture many routing problems arising in heterogeneous cyber-physical systems~\cite{AB:95:SIAMDM,BC:00:IPL,A:01:NETWORKS,B:02:TCS,M:15:IPL,FHPS:06:ALGOR,B:08:JDALG,CHHK:22:ALGCA,CCHKK:20:ICS,KM:2018:incollection}. 
Modern transportation, manufacturing, communication, and robotic systems are often composed of multiple operational regions with different physical environments, hardware configurations, or communication protocols. 
Within each individual region, transportation or communication costs often satisfy the triangle inequality, as the underlying infrastructure is relatively homogeneous~\cite{SHES:16:IJCAI,SYKK:17:ijcai,ZXPX:25:aaai,ZXPX:2026:article}. 
However, moving between different regions may incur additional overhead caused by equipment reconfiguration, protocol conversion, synchronization delays, security constraints, or other operational factors~\cite{SNK:14:walcom,SNK:2016:article,MRM:2017:article,MRM:2019:article}. 
Thus, although routing costs within each region remain metric, the overall cost function may violate the triangle inequality.

This combination of heterogeneous regional structures and metric intra-region costs arises in various real-world applications.
In intelligent manufacturing systems such as \acl{PCB} assembly, switching between different production modules requires additional setup operations beyond the physical travel distance~\cite{MRM:2017:article}. 
Similar situations occur in multi-stage logistics systems, heterogeneous robotic platforms, cloud-edge collaborative computing, and hierarchical communication networks~\cite{TDZ:04:ICRA,GGMSTZ:10:IEEE_J_EPM,ZLG:22:ISAAC,BEX:24:arXiv,ZZG:25:COCOON,ZSX:26:AAAI}, where traversing different operational domains introduces transition costs that cannot be explained solely by geometric distance. These scenarios motivate optimization models that preserve metric structures locally while explicitly incorporating biased costs across different regions.

To model such routing problems, the \aca{BTSP} (\ac{BTSP}) was introduced in~\cite{MRM:2017:article}.
The problem is defined on an underlying complete metric graph whose vertex set is partitioned into $k$ disjoint regions. 
Let $c\lrA{\cdot,\cdot}$ denote the original metric edge cost. 
For any edge $xy$ connecting vertices from different regions, its biased cost is defined as $\c\lrA{x,y}=\beta\lrA{x,y}c\lrA{x,y}$, where $\beta\lrA{x,y} \ge 1$ is a symmetric bias factor.
Edges within the same region keep their original costs. 
Thus, each region preserves the metric structure, while inter-region transitions incur additional biased costs.
Its path counterpart, the \acl{BTSPP} (\aca{BTSPP}), asks for a minimum-cost Hamiltonian path between two given vertices~\cite{MRM:2019:article}.

The main algorithmic challenge of biased routing problems lies in exploiting local metric structures while handling the absence of a global triangle inequality and the potentially large costs of inter-region transitions.
Unlike the metric \ac{TSP}, where the triangle inequality holds globally, the \ac{BTSP} satisfies this property only locally. 
Consequently, many classical approximation techniques cannot be applied directly. 
For example, the work~\cite{MRM:2017:article} showed that a direct adaptation of the classical $3/2$-approximation algorithm~\cite{C:2022:article} for the metric \ac{TSP} does not yield a constant-factor approximation guarantee. This limitation calls for new algorithmic techniques that can combine local metric structures with biased inter-region connections.

Existing approximation results remain limited.
To date, approximation algorithms have only been established for the bipartite case ($k=2$), where a $7/2$-approximation algorithm for the \ac{BTSP} was developed in~\cite{MRM:2017:article}, and a $4$-approximation algorithm for the \ac{BTSPP} was proposed in~\cite{MRM:2019:article}. 
However, these techniques rely on structural properties specific to the bipartite case and do not extend naturally to larger numbers of regions. 
Whether constant-factor approximation algorithms exist for fixed $k\ge3$ was posed as an open question in~\cite{MRM:2017:article} for the \ac{BTSP} and in~\cite{MRM:2019:article} for the \ac{BTSPP}.

\subsection{Related Work}

The \ac{BTSP} combines metric routing structures within individual regions with biased costs between different regions. 
Therefore, its approximation algorithms are closely related to both classical metric routing problems and various models that relax or modify the triangle inequality. 
We first review approximation results for metric routing problems, which provide the fundamental algorithmic tools used in our framework, and then discuss representative non-metric \ac{TSP} variants.

\subsubsection{Metric Variants}

The metric \ac{TSP} and the metric \ac{TSPP} are among the most extensively studied routing problems and provide important algorithmic foundations for many structured routing problems.
For the metric \ac{TSP}, a classical $3/2$-approximation algorithm~\cite{C:2022:article} remained the best-known guarantee for several decades before being improved to $\alpha<3/2-10^{-36}$~\cite{KKG:21:stoc,KKG:23:ipco}. Subsequent work further improved the approximation ratio to $\alpha<3/2-10^{-34}$~\cite{LNJ:24:ICALP}.
The metric \ac{TSPP}, which asks for a minimum-cost Hamiltonian path between two specified vertices, has also been studied extensively. Early approximation algorithms achieved an approximation ratio of $5/3$~\cite{H:1991:article}, which was later improved to $3/2$~\cite{Z:19:soda}, and subsequently to $\alpha+\varepsilon$ for any fixed $\varepsilon>0$~\cite{KKG:21:stoc}.
These approximation algorithms have become fundamental building blocks for various routing problems where metric structures are preserved on appropriate components of the problem instance.

\subsubsection{Models for Non-Metric \acp{TSP}}

A large body of research has investigated variants of the \ac{TSP} where the triangle inequality may not hold.
One representative model is the $\tau$-triangle inequality, which requires
$c\lrA{x,y}\leq \tau\lra{c\lrA{x,z}+c\lrA{z,y}}$
for every triple of vertices.
A variety of approximation results have been established under this relaxed assumption, with guarantees depending on the value of $\tau$~\cite{M:15:IPL,AB:95:SIAMDM,BC:00:IPL,A:01:NETWORKS,B:02:TCS,FHPS:06:ALGOR}. Similar ideas have also been applied to other network design problems~\cite{KM:2018:incollection,CCHKK:20:ICS,CHHK:22:ALGCA,B:08:JDALG}.

Another line of research considers routing problems with additional structural constraints. Representative examples include the \ac{SGPP} and its special case, the \ac{SPP}, where vertices belonging to the same subgroup are required to be visited consecutively~\cite{TDZ:04:ICRA,GGMSTZ:10:IEEE_J_EPM,SHES:16:IJCAI,SYKK:17:ijcai}. Recent work has further improved approximation algorithms for these problems by exploiting parameters such as the subgroup size or the total number of subgroups~\cite{ZXPX:25:aaai,ZXPX:2026:article}.
More recently, parameterized approaches have studied nearly metric graphs by introducing quantitative measures of how much the triangle inequality is violated~\cite{ZLG:22:ISAAC,BEX:24:arXiv,ZZG:25:COCOON,ZSX:26:AAAI}. These parameters include the number of vertices involved in violated triangle inequalities and the minimum number of vertices whose deletion makes the graph metric. 
These parameters have led to various 
parameterized approximation algorithms for nearly metric graphs.

The \ac{BTSP} studied in this paper adopts a different modeling perspective by partitioning the vertex set into a fixed number of metric regions, where intra-region edges preserve the metric structure while inter-region edges may have their costs multiplied by arbitrary factors no smaller than one. 
Consequently, the graph may contain an unbounded number of violations of the triangle inequality across regions, making existing techniques developed for nearly metric graphs and subgroup-constrained routing problems difficult to extend.

\subsection{Our Contributions}

In this work, we develop the first constant-factor approximation framework for the \ac{BTSP} and the \ac{BTSPP} with an arbitrary fixed number of regions. 
Our results resolve the open questions regarding the existence of constant-factor approximation algorithms beyond the bipartite case~\cite{MRM:2017:article,MRM:2019:article}. 
In addition, for the bipartite case, we improve the approximation guarantee for the \ac{BTSPP}. 
Our main contributions are summarized as follows.

\subsubsection{An $O\lrA{k}$-Approximation Framework for Any Fixed $k$}

We establish a general framework that extends biased routing problems from the bipartite case to any fixed number of metric regions.
In particular, we present an $\lrA{\alpha k+1}$-approximation algorithm for the \ac{BTSP} and a $\lrA{2k+1}$-approximation algorithm for the \ac{BTSPP}.
Both algorithms run in time $N^{O\lra{2^kk!}}$, which is polynomial for any fixed $k$.

Existing approximation algorithms~\cite{MRM:2017:article,MRM:2019:article} for the bipartite case are based on minimum spanning trees and rely on detailed case analyses to handle different configurations arising from interactions between the two regions.
Such techniques become difficult to extend when the number of regions grows, since the structural configurations induced by repeated visits to multiple regions become increasingly complicated.
In contrast, our method adopts a fundamentally different perspective by showing that the state sequence of an optimal solution can be compressed into a bounded representation whose length depends only on $k$.

The main technical contribution is a new structural characterization of optimal solutions based on state compression.
We represent a Hamiltonian path by a sequence of states, where each state describes a subpath whose endpoints belong to the same region.
Although the state sequence of an optimal solution may contain a number of states depending on the instance size, we prove that it can be transformed into an irreducible state sequence through a sequence of legal merge operations without increasing the solution cost. The resulting irreducible sequence has length bounded only by the number of regions, specifically $O\lra{2^k k!}$.
Moreover, the compressed sequence preserves sufficient information for solution reconstruction: for each region, either all of its original states are retained, or at least one representative subpath within that region is preserved.
This bounded representation allows us to enumerate all possible irreducible state sequences in polynomial time for any fixed $k$.
For each enumerated state sequence, we reconstruct a complete solution by identifying the vertices removed during compression and recovering the corresponding subpaths within individual regions using approximation algorithms.
Since each individual region preserves the metric structure, classical approximation algorithms for the metric \ac{TSP} and its path counterpart \aca{TSPP} can be employed to construct the required intra-region cycles or paths.
By patching these intra-region parts into the enumerated structure, we obtain feasible solutions with provable approximation ratios.

The proposed framework provides a general methodology for routing problems that preserve metric structures within individual regions while allowing biased costs between different regions.
Unlike previous approaches that rely on case-specific analyses for the bipartite case, our method captures the essential structural properties of optimal biased routing solutions and extends naturally to any fixed number of regions.

\subsubsection{An Improved $\lrA{\alpha+2+\varepsilon}$-Approximation Algorithm for the \ac{BTSPP} with $k=2$}

We focus on the classical bipartite case of the \ac{BTSPP}, and present a polynomial-time $\lrA{\alpha+2+\varepsilon}$-approximation algorithm for any constant $\varepsilon>0$, improving the previous best-known $4$-approximation algorithm~\cite{MRM:2019:article}.

The algorithm builds upon the state-compression framework developed above and further exploits structural properties specific to the bipartite case. We first classify the state sequence of an optimal Hamiltonian path into several structural cases according to the arrangement of consecutive states from the two regions. For each case, we identify a compressed state sequence of constant length through legal compression operations while preserving sufficient information for reconstructing a feasible Hamiltonian path. The subpaths represented by the compressed states are then reconstructed independently within each metric region using a simple $2$-approximation algorithm for the metric \ac{TSPP}.

The most challenging situation arises when the state sequence alternates frequently between the two regions, making direct compression ineffective. To address this difficulty, we establish an averaging argument showing that there exists a transition between two consecutive visits to the same region whose cost is at most a small fraction of the optimal cost. This enables an additional patching strategy whose extra cost can be made arbitrarily small by choosing an appropriate constant threshold. Combining the reconstruction procedures and analyses for all structural cases yields the overall approximation ratio of $\alpha+2+\varepsilon$ while keeping polynomial running time.

\subsection{Organization of the Paper}

The remainder of the paper is organized as follows.
Section~\ref{sec:2} introduces preliminaries of $k$-partitioned networks, detailing the biased edge-cost function and its structural properties.
Section~\ref{sec:3} presents our generalized $O\lrA{k}$-approximation algorithms for both the \ac{BTSP} and the \ac{BTSPP}, including the state-compression framework and enumeration strategy for arbitrary fixed $k$.
Section~\ref{sec:4} presents our improved $\lrA{\alpha+2+\varepsilon}$-approximation algorithm for the bipartite case of the \ac{BTSPP}, based on refined state-compression, structural case analysis, and an averaging argument.
Section~\ref{sec:5} concludes the paper.

\section{Preliminaries}\label{sec:2}

We first introduce the notation and structural properties of $k$-partitioned networks used throughout the paper.

For any positive integer $t$, let $\lrB{t}$ denote the set of integers $\lrc{1,\ldots,t}$.
Consider a graph $\G=\lrA{\V,\E}$ with a cost function $c$.
For any (multi-)graph $\G'$ whose vertex set is a subset of $\V$, let $\V\lra{\G'}$ and $\E\lra{\G'}$ denote its vertex set and edge (multi-)set, respectively.
For any edge (multi-)set $\F\subseteq \E\lra{\G'}$, we define its cost as $c\lrA{\F}=\sum_{e\in \F} c\lrA{e}$.

A walk $\pi=\lrd{v_1,v_2,\ldots,v_\ell}$ in $\G$ is a sequence of vertices where consecutive vertices are connected by edges.
The vertex set and edge (multi-)set of $\pi$ are denoted by $\V\lrA{\pi}=\lrc{v_1,v_2,\ldots,v_\ell}$ and $\E\lrA{\pi}=\lrc{v_1v_2,v_2v_3,\ldots,v_{\ell-1}v_\ell}$, respectively.
The cost of $\pi$ is $c\lrA{\pi}=c\lrA{\E\lrA{\pi}}$.
A walk without repeated vertices is called a path. A walk $\lrd{v_1,v_2,\ldots,v_\ell,v_1}$ is called a cycle if $v_1,\ldots,v_\ell$ are pairwise distinct.
A Hamiltonian path (cycle) in $\G$ is a path (cycle) that contains every vertex of $\G$.
For a path $\pi=\lrd{v_1,\ldots,v_\ell}$, the vertices $v_1$ and $v_\ell$ are called its terminals.
Given a walk $\pi=\lrd{\ldots,x,y,z,\ldots}$, replacing the segment
$\lrd{x,y,z}$ by $\lrd {x,z}$ is called shortcutting vertex $y$.
When the underlying edge-cost function satisfies the triangle inequality, shortcutting does not increase the total cost.

\subsection{$k$-Partitioned Networks}

We consider a complete undirected graph
$\G=\lrA{\V,\E}$ with a metric edge-cost function
$c:\E\rightarrow\mathbb{R}_{\geq0}$ satisfying $c\lrA{x,x}=0$, $c\lrA{x,y}=c\lrA{y,x}$, and $c\lrA{x,y} \leq c\lrA{x,z} + c\lrA{z,y}$ (the triangle inequality) for all $x,y,z\in \V$.
The set $\V$ consists of $N$ vertices and is partitioned into $k$ non-empty regions $\lrc{\V_1,\V_2,\ldots,\V_k}$,
where $\bigcup_{i\in\lrB{k}}\V_i=\V$, and $\V_i\cap \V_j=\emptyset$ for any $i\neq j$.
For every edge connecting two different regions, its cost may incur an additional transition overhead.
In particular, let $\beta$ be a symmetric multiplicative bias factor defined for every pair of vertices from different regions, satisfying $\beta\lrA{u,v}=\beta\lrA{v,u}\ge 1$.
The biased edge-cost function
$\c:\E\rightarrow\mathbb{R}_{\geq0}$ is defined as
\[
\c\lrA{u,v}=
\begin{cases}
c\lrA{u,v}, & u,v\in \V_i\text{ for some }i\in\lrB{k},\\
\beta\lrA{u,v} c\lrA{u,v}, & u\in \V_i,\ v\in \V_j,\ i\neq j.
\end{cases}
\]

Under this biased cost function, each individual region preserves the original metric structure, while transitions between different regions may incur additional multiplicative costs.
Thus, for every $i\in\lrB{k}$, the induced subgraph $\G\lrb{\V_i}$ remains a metric graph with respect to $\c$.

Although $\c$ does not necessarily satisfy the triangle inequality over the entire graph, it preserves this property for several important configurations.
For any vertices $x,y,z\in \V$, 
\[
\c\lrA{x,y}+\c\lrA{y,z}\geq \c\lrA{x,z}
\]
whenever either $x,y,z\in \V_i$ for some $i\in\lrB{k}$, or $x,z\in \V_i$ and $y\notin \V_i$.
The first case follows from the metricity of each region.
For the second case, $\c\lrA{x,z}=c\lrA{x,z}$ because $x$ and $z$ belong to the same region, while $\c\lrA{x,y}\geq c\lrA{x,y}$ and $\c\lrA{y,z}\geq c\lrA{y,z}$ by definition.
More generally, shortcutting all internal vertices in a sequence $\lrd{x,\ldots,z}$, where $x, z \in \V_i$, also preserves the biased triangle inequality.

Unless otherwise stated, all costs in the remainder of this paper are evaluated according to the biased cost function $\c$.

The \ac{BTSP} seeks a minimum-cost Hamiltonian cycle under the biased cost function $\c$.
The \ac{BTSPP} asks for a minimum-cost Hamiltonian path between two specified terminals $s$ and $t$, where $s$ and $t$ are not necessarily in the same region.

For each $\V_i$, let $\sigma^*_{\V_i}$ denote an optimal Hamiltonian cycle in the metric subgraph $\G\lrb{\V_i}$, and let $\pi^*_{\V_i}$ denote an optimal Hamiltonian path in $\G\lrb{\V_i}$ between any two vertices of $\V_i$.
Let $\sigma^*$ and $\pi^*$ denote optimal solutions to the \ac{BTSP} and the \ac{BTSPP}, respectively.
The following property will be frequently used in our analysis.

\begin{lemma}\label{bound}
It holds that $\c\lra{\sigma^*_{\V_i}}\leq \c\lra{\sigma^*}$ and $\c\lra{\pi^*_{\V_i}}\leq \c\lra{\pi^*}$ for all $i\in\lrB{k}$.
\end{lemma}
\begin{proof}
Fix any $i\in\lrB{k}$. 
Let $v_1,\ldots,v_{\size{\V_i}}$ be the vertices of $\V_i$ in the order in which they occur in $\sigma^*$. For each $j\in\lrB{\size{\V_i}}$, shortcut all internal vertices in the segment of $\sigma^*$ from $v_j$ to $v_{j+1}$, where $v_{\size{\V_i}+1}=v_1$. 
The resulting cycle is a Hamiltonian cycle in $\G\lrb{\V_i}$. By the biased triangle inequality, these shortcuts do not increase the cost. Hence, $\c\lra{\sigma^*_{\V_i}}\leq \c\lra{\sigma^*}$.
Now consider $\pi^*$. Let $u_1,\ldots,u_{\size{\V_i}}$ be the vertices of $\V_i$ in the order in which they occur in $\pi^*$. The subpath of $\pi^*$ from $u_1$ to $u_{\size{\V_i}}$ contains all vertices of $\V_i$. For each $j\in\lrB{\size{\V_i}-1}$, shortcut all internal vertices in the segment of this subpath from $u_j$ to $u_{j+1}$. The resulting path is a Hamiltonian path in $\G\lrb{\V_i}$ with terminals $u_1$ and $u_{\size{\V_i}}$. By the biased triangle inequality, these shortcuts do not increase the cost. Moreover, the subpath from $u_1$ to $u_{\size{\V_i}}$ has no greater cost than $\pi^*$ itself. Since $\pi^*_{\V_i}$ is an optimal Hamiltonian path in $\G\lrb{\V_i}$ over all choices of its terminals, $\c\lra{\pi^*_{\V_i}} \leq \c\lra{\pi^*}$.
\end{proof}

\section{Structural Properties of the \ac{BTSPP}}\label{sec:3}
In this section, we analyze the structural properties of the \ac{BTSPP}, which is also useful for the \ac{BTSP}, as deleting any edge from a Hamiltonian cycle yields a Hamiltonian path.

We abstract any path in $\G$ into a sequence of \emph{states}, where each state denotes a specific vertex or an edge whose terminals belong to the same region $\V_i$ for some $i\in\lrB{k}$.

\begin{definition}[Trivial and Non-trivial States]\label{def1}
A state is denoted by a tuple $\lra{\V_i, u, v}$, where $u, v \in \V_i$, which is called \emph{trivial} if $u = v$, and \emph{non-trivial} otherwise.
\end{definition}

Given a path $\pi$ in $\G$, we construct its initial state sequence $\zeta_\pi$ by traversing $\pi$ from one terminal to the other, recording a trivial state for each vertex in $\pi$.
For instance, suppose $\V=\V_1\cup \V_2\cup \V_3$, i.e., $k=3$, where $\V_1=\lrc{u_1,u_2,u_3,u_4}$, $\V_2=\lrc{v_1,v_2}$, and $\V_3=\lrc{w_1,w_2,w_3}$.
Let a path in $\G$ be 
\begin{equation}\label{sig1}
\pi =\lrd{u_1,u_2,v_1,u_3,w_1,v_2,w_2,u_4,w_3}.
\end{equation}
The state sequence of $\pi$ is then defined as
\begin{equation}\label{zeta1}
\begin{split}
\zeta_\pi&=\left\langle\lra{\V_1,u_1,u_1},\lra{\V_1,u_2,u_2},\lra{\V_2,v_1,v_1},\lra{\V_1,u_3,u_3},\right. \\
&\quad\quad\lra{\V_3,w_1,w_1},\lra{\V_2,v_2,v_2},\lra{\V_3,w_2,w_2},\\
&\quad\quad\left.\lra{\V_1,u_4,u_4},\lra{\V_3,w_3,w_3}\right\rangle.
\end{split}
\end{equation}

Clearly, every state in $\zeta_\pi$ is trivial.
The length of $\zeta_\pi$, denoted by $\size{\zeta_\pi}$, is defined as the total number of states it contains.
Hence, $\size{\zeta_\pi}=9$, which also equals the total number of vertices $N$ in $\V$, as $\pi$ is a Hamiltonian path in $\G$ in this example.

In the following, we only consider state sequences that correspond to a path in $\G$.
Clearly, such a state sequence can be used to reconstruct the corresponding path in $\G$.
For any state sequence $\zeta$, we use $\P_\zeta$ to denote the corresponding path in $\G$. Moreover, for any $u,v\in \V\lra{\P_\zeta}$, we use $\P_\zeta\lrA{u,v}$ to denote the subpath between $u$ and $v$ in $\P_\zeta$.

Our goal is to use a state sequence to reconstruct an approximate Hamiltonian path in $\G$.
Since the state sequence of an optimal Hamiltonian cycle in $\G$ has length $N$, it cannot be enumerated in polynomial time, and thus cannot be used directly to reconstruct a Hamiltonian path in $\G$.
To ensure the efficiency of the reconstruction process, we aim to reduce the length of the state sequence to a constant that depends solely on $k$, independent of $N$.
Specifically, we iteratively compress the state sequence by merging consecutive states into a single new state.
This abstraction corresponds to shortcutting the internal vertices of the corresponding subpath in $\G$.
Based on the definitions of the \ac{BTSPP} and states, we introduce two \acp{MO} for compressing a \emph{contiguous} state sequence: \ac{MO}-1 and \ac{MO}-2.

\begin{definition}[\ac{MO}-1]\label{def2}
Given a state sequence $\zeta$, any contiguous subsequence of $\zeta$ can be merged into a single state if its first and last states are associated with the same $\V_i$ for some $i\in\lrB{k}$.
Specifically, suppose $\zeta=\lrd{\ldots,\lra{\V_i,v,x},\ldots,\lra{\V_i,y,u},\ldots}$ contains a contiguous subsequence starting with $\lra{\V_i,v,x}$ and ending with $\lra{\V_i,y,u}$.
Then, after applying the merge operation \ac{MO}-1 to this subsequence, the resulting sequence is $\zeta' =\lrd{\ldots,\lra{\V_i,v,u},\ldots}$.
\end{definition}

In this context, \ac{MO}-1 corresponds to shortcutting all internal vertices on the subpath $\P_{\zeta}\lrA{v,u}$ of the path $\P_{\zeta}$.
Since $u,v\in \V_i$, by the biased triangle inequality, the cost of $\P_{\zeta}\lrA{v,u}$ is at least $\c\lrA{u,v}$. 
Therefore, the cost of $\P_{\zeta'}$ is guaranteed to be no greater than that of $\P_{\zeta}$.
Moreover, since both $\P_{\zeta}$ and $\P_{\zeta'}$ are paths in $\G$, the resulting state $\lra{\V_i,v,u}$ must be non-trivial.

In the following discussion, whenever we refer to a subsequence of a sequence, we mean a {contiguous} subsequence.

While \ac{MO}-1 merges only one subsequence at a time, multiple independent merges can be performed simultaneously.
\begin{definition}[\ac{MO}-2]\label{def3}
Given a state sequence $\zeta$, one can simultaneously apply \ac{MO}-1 to merge multiple subsequences of $\zeta$ if these subsequences are disjoint.
\end{definition}

For example, \ac{MO}-2 can be applied to simultaneously merge the following two disjoint subsequences of $\zeta_\pi$ from \eqref{sig1}:
\[
\lrd{\lra{\V_1,u_1,u_1},\lra{\V_1,u_2,u_2},\lra{\V_2,v_1,v_1},\lra{\V_1,u_3,u_3}}
\]
and 
\[
\lrd{\lra{\V_3,w_1,w_1},\lra{\V_2,v_2,v_2},\lra{\V_3,w_2,w_2}},
\] 
yielding a shortened sequence of length $4$:
\begin{equation}\label{exp1}
\lrd{\lra{\V_1,u_1,u_3},\lra{\V_3,w_1,w_2},\lra{\V_1,u_4,u_4},\lra{\V_3,w_3,w_3}}.
\end{equation}

Note that the resulting sequence in \eqref{exp1}  no longer contains any state associated with $\V_2$, thus completely losing the routing information of the vertices in $\V_2$.
To prevent such loss of information and ensure the feasibility of reconstructing an approximate Hamiltonian path in $\G$, we define the following condition for the \emph{legal} compression of state sequences.

\begin{definition}[Compression Condition]\label{def4}
Let $\zeta$ be a state sequence, and let $\mathscr{V}_\zeta=\lrc{\V_{i_1}, \ldots, \V_{i_\ell}}$ denote the set of regions appearing in $\zeta$.
For each $j \in\lrB{\ell}$, let $N_{i_j}$ denote the number of states associated with $\V_{i_j}$ in $\zeta$.
The sequence $\zeta$ can be legally compressed into a shorter sequence $\zeta'$ using \ac{MO}-1 or \ac{MO}-2 if, in $\zeta'$, for every $j\in\lrB{\ell}$, one of the following holds:
\begin{enumerate}
    \item $\zeta'$ contains at least one non-trivial state associated with $\V_{i_j}$, or
    \item $\zeta'$ contains $N_{i_j}$ states associated with $\V_{i_j}$ (i.e., every state associated with $\V_{i_j}$ in $\zeta$ was left completely unmerged).
\end{enumerate}
\end{definition}

The state sequence $\zeta$ of a Hamiltonian path in $\G$ satisfies $\mathscr{V}_\zeta=\lrc{\V_1,\ldots, \V_k}$ and $\sum_{i=1}^k N_i = N$, where $N_i=\size{\V_i}$ is the number of states associated with $\V_i$.
Moreover, any state sequence $\zeta'$ obtained from $\zeta$ through legal compressions satisfies $\mathscr{V}_{\zeta'}=\mathscr{V}_{\zeta}$, and for each $\V_i\in\mathscr{V}_{\zeta'}$, $\zeta'$ either contains a non-trivial state associated with $\V_i$ or contains all states associated with $\V_i$ in $\zeta$.

Therefore, a valid legal compression is a local operation that modifies only the states corresponding to a non-empty subset $\mathscr{V}'_\zeta\subseteq\mathscr{V}_\zeta$, ensuring that for each $\V_i\in\mathscr{V}'_\zeta$ there exists at least one non-trivial state associated with $\V_i$, while leaving all states associated with every $\V_j\in\mathscr{V}_\zeta\setminus\mathscr{V}'_\zeta$ completely unchanged.
It satisfies the following inherited property.

\begin{property}\label{inhp}
Given a state sequence $\zeta$, for any subsequence $\zeta'$ of $\zeta$, if $\zeta'$ can be legally compressed into a shorter sequence, then $\zeta$ can also be legally compressed into a shorter sequence by applying the same compression to $\zeta'$.
\end{property}

Any state sequence is called an \emph{irreducible} sequence if it cannot be further legally compressed, and \emph{reducible} otherwise.

Given the initial state sequence $\zeta_{\pi}$ of a Hamiltonian path $\pi$ in $\G$, we know that the length of $\zeta_{\pi}$ is exactly $N$.
Thus, a final irreducible sequence $\zeta$ can be obtained by iteratively applying legal compressions at most $O\lrA{N}$ times. 
Note that a state sequence may admit multiple legal compressions, and different choices may lead to different irreducible sequences.
Our goal is to prove that the length of any irreducible sequence is bounded by a constant that depends solely on $k$, independent of the total number of vertices $N$.
 
To this end, we represent any state sequence $\zeta$, as a string $\S_\zeta$ over the alphabet $\Sigma \coloneqq \mathscr{V}_\zeta$.
A subsequence of $\S_\zeta$ is called a \emph{closed sub-chain} if it starts and ends with the same character $\V_i$ for some $\V_i\in\Sigma$ and has length at least 2; in this case, we say that $\V_i$ delimits the closed sub-chain.

For example, using $\zeta_\pi$ from \eqref{zeta1}, we obtain $\Sigma =\lrc{\V_1, \V_2, \V_3}$ and $\S_{\zeta_\pi} =\lrd{\V_1,\V_1,\V_2,\V_1,\V_3,\V_2,\V_3,\V_1,\V_3}$.
Moreover, we observe that in $\S_{\zeta_\pi}$, $\lrd{\V_1,\V_1}$ is a closed sub-chain delimited by $\V_1$ and $\lrd{\V_3,\V_2,\V_3}$ is a closed sub-chain delimited by $\V_3$.

\begin{lemma}[Safe Compressible Window] \label{lemma:window}
Let $\WW$ be a subsequence of $\S_\zeta$, referred to as a window, and let $\Sigma_\WW \subseteq \Sigma$ be the set of characters appearing in $\WW$.
If $\WW$ contains $\size{\Sigma_\WW}$ pairwise disjoint closed sub-chains such that each character in $\Sigma_\WW$ delimits one of these closed sub-chains, then $\zeta$ can be legally compressed.
\end{lemma}
\begin{IEEEproof}
Since $\WW$ contains $\size{\Sigma_\WW}$ pairwise disjoint closed sub-chains such that each character in $\Sigma_\WW$ delimits one of these closed sub-chains, we associate each character in $\Sigma_\WW$ with a distinct closed sub-chain it delimits.
We then perform a compression on $\zeta$ by applying \ac{MO}-2 to the $\size{\Sigma_\WW}$ subsequences corresponding to the $\size{\Sigma_\WW}$ disjoint closed sub-chains.

Denote the resulting sequence after this compression as $\zeta'$, and its corresponding string as $\S_{\zeta'}$.

On one hand, for each $\V_i \in \Sigma_\WW$, merging the subsequence corresponding to the closed sub-chain that $\V_i$ delimits produces a non-trivial state associated with $\V_i$ by Definitions \ref{def2} and \ref{def3}.
Thus, $\zeta'$ satisfies property 1) in Definition \ref{def4} for each $\V_i\in \Sigma_\WW$.

On the other hand, for each $\V_i\in\Sigma\setminus \Sigma_\WW$, its states remain unchanged during this localized compression.
Consequently, if $\V_i$ has a non-trivial state in $\zeta$, it retains one in $\zeta'$ (satisfying Condition 1). If all its states in $\zeta$ are trivial, their exact number is preserved in $\zeta'$ (satisfying Condition 2).

Therefore, the compression is legal.
\end{IEEEproof}

To prove the length bound of any irreducible sequence, we first introduce the concept of the maximum gap.

\begin{definition}[Maximum Gap] \label{def:maxgap}
For a sequence $\S_\zeta$ of length $L$ and a character $\V_i \in \Sigma$, the maximum gap $\Delta\lra{\V_i}$ is the maximum number of consecutive characters in $\S_\zeta$ that do not contain $\V_i$. It is the maximum among:
\begin{enumerate}
    \item The number of characters strictly preceding the first occurrence of $\V_i$,
    \item The number of characters strictly between two adjacent occurrences of $\V_i$, and
    \item The number of characters strictly following the last occurrence of $\V_i$.
\end{enumerate}
\end{definition}

Note that $\Delta\lra{\V_i}$ quantifies the largest contiguous segment of $\S_\zeta$ that completely lacks the character $\V_i$.

\begin{lemma}[Structural Property]\label{lem:bound}
For any $k \geq 1$, any irreducible state sequence $\zeta$ whose string $\S_\zeta$ is over an alphabet of size $k$ has length at most $f\lrA{k}$, where $f\lrA{k} = 2k  f\lrA{k-1} + 2k$ and $f\lrA{1}=1$. In particular, $f\lrA{k} = O\lra{2^k k!}$.
\end{lemma}
\begin{IEEEproof}
The proof is by induction on the alphabet size $k$.

For the base case where $k=1$, any two adjacent characters in $\S_\zeta$ must be distinct; otherwise, one could apply the \ac{MO}-1 operation to compress the state sequence $\zeta$.
Thus, the maximum length of any irreducible state sequence over a single-character alphabet is exactly $1$. Since $f\lrA{1} = 1$ by definition, the lemma holds for the base case $k=1$.

Assume the lemma holds for all alphabet sizes up to $k-1$. 
That is, for any $i \in\lrB{k-1}$, any irreducible state sequence whose string is over an alphabet of size $i$ has a maximum length of $f\lrA{i}$.
We then consider an irreducible state sequence $\zeta$ whose string $\S_\zeta$ is over an alphabet $\Sigma$ of $k$ characters.
We analyze the sequence according to the maximum gap defined in Definition~\ref{def:maxgap}.

\paragraph{Case~1: Existence of a large gap} Suppose there exists at least one character $\V_i \in \Sigma$ such that its maximum gap satisfies $\Delta\lra{\V_i}> f\lrA{k-1}$.
This implies that there is a subsequence $\S'$ within $\S_\zeta$ of length greater than $f\lrA{k-1}$ that completely lacks the character $\V_i$. Thus, the subsequence $\S'$ is constructed over an alphabet of at most $k-1$ characters.
By our inductive hypothesis, any sequence over at most $k-1$ characters with a length exceeding $f\lrA{k-1}$ must be reducible. 
Thus, there exists a legal compression localized within $\S'$. 
By Property~\ref{inhp}, this local legal compression yields a legal compression of the global sequence $\zeta$, directly contradicting our initial assumption that $\zeta$ is irreducible.

\paragraph{Case~2: Absence of large gaps} If Case 1 does not hold, then for every character $\V_i \in \Sigma$, its maximum gap satisfies $\Delta\lra{\V_i} \le f\lrA{k-1}$. Hence, whenever a character appears again after the current position, its next occurrence is encountered within at most $f\lrA{k-1}+1$ scanned positions.

We greedily scan $\S_\zeta$ from left to right to construct closed sub-chains for $\V_1, \dots, \V_k$ sequentially. 
For any character $\V_i$, locating its first occurrence after the current scan index (if it exists) requires at most $f\lrA{k-1}+1$ scanned positions, and locating its second occurrence (if it exists) requires at most another $f\lrA{k-1}+1$ positions. Thus, whenever a closed sub-chain for $\V_i$ can be formed, it occupies at most $2f\lrA{k-1}+2$ scanned positions.
We consider two possibilities for this scan:
\begin{itemize}
    \item If the construction succeeds for all $k$ characters, the constructed sub-chains together form a window $\WW$ of length at most
    \begin{equation}
    L_{\max} = 2k f\lrA{k-1} + 2k.
    \end{equation}
    Since $\WW$ contains pairwise disjoint closed sub-chains for all $k$ characters in $\Sigma$, Lemma~\ref{lemma:window} implies that $\zeta$ can be legally compressed, contradicting the assumption that $\zeta$ is irreducible.
    
    \item Otherwise, the construction fails at some character $\V_i$ because $\V_i$ appears at most once in the remaining suffix of $\S_\zeta$. Since $\Delta\lra{\V_i} \le f\lrA{k-1}$, the length of this remaining suffix is at most $2f\lrA{k-1}+1$. Consequently, the total length of $\S_\zeta$ cannot exceed $L_{\max}$, directly yielding $\size{\S_\zeta} \le L_{\max}$.
\end{itemize}

Therefore, to avoid contradiction, any irreducible state sequence $\zeta$ over an alphabet of size $k$ must satisfy $\size{\S_\zeta} \le L_{\max} = 2k f\lrA{k-1} +2k$.

It can be verified that the recursive bounding function satisfies $f\lrA{k} = {O}\lra{2^k k!}$. 
Consequently, the maximum length of any irreducible state sequence over an alphabet of size $k$ is bounded by ${O}\lra{2^k k!}$, which completes the proof.
\end{IEEEproof}

\section{Approximation Algorithms for the \ac{BTSP} and the \ac{BTSPP}}
\label{sec:algorithms}

In this section, we use the structural properties established in the previous section to design approximation algorithms for the \ac{BTSP} and the \ac{BTSPP}. Specifically, we develop an $\lrA{\alpha k + 1}$-approximation algorithm for the \ac{BTSP} and a $\lrA{2k + 1}$-approximation algorithm for the \ac{BTSPP}, both running in polynomial time for any fixed constant $k$.

\subsection{Algorithm Overview}

At a high level, our algorithms leverage the bounded length of irreducible state sequences. By Lemma~\ref{lem:bound}, the length of any irreducible state sequence $\zeta$ is bounded by a constant $f\lrA{k} = O\lra{2^k k!}$, which depends solely on the number of regions $k$. Consequently, the total number of paths $\P_\zeta$ in $\G$ corresponding to all possible irreducible state sequences is bounded by $N^{O\lrA{f\lrA{k}}}$. This allows us to enumerate all such paths in polynomial time.

For each enumerated path $\P_\zeta$, some vertices in $\G$ may remain uncovered. Let $\overline{\V}_i \coloneqq \V_i \setminus \V\lra{\P_\zeta}$ denote the set of uncovered vertices in region $\V_i$. 
According to Definition~\ref{def4}, if $\overline{\V}_i \neq \emptyset$, the state sequence $\zeta$ could not have left all states associated with $\V_i$ completely unmerged. Thus, $\zeta$ must contain at least one non-trivial state $\lra{\V_i, u_i, v_i}$. In the corresponding path $\P_\zeta$, this state represents the compressed subpath and directly corresponds to the edge $u_iv_i$.

To cover the vertices in $\overline{\V}_i$, we compute a Hamiltonian cycle $\CC_i$ in the induced subgraph $\G\lrb{\overline{\V}_i \cup\lrc{u_i, v_i}}$. We then extract a Hamiltonian path $\P_i$ between $u_i$ and $v_i$ from $\CC_i$, and patch it into $\P_\zeta$ by replacing the edge $u_iv_i$. By applying this patching operation across all regions where $\overline{\V}_i \neq \emptyset$, we reconstruct a Hamiltonian path or cycle in $\G$.

\subsection{A $\lrA{2k+1}$-Approximation Algorithm for the \ac{BTSPP}}

For the \ac{BTSPP}, our goal is to find a minimum-cost Hamiltonian path between two given vertices $s$ and $t$. 
We instantiate the general framework by using the standard double-tree algorithm~\cite{WS:2011:book}, which yields an approximation ratio of 2 for the \ac{TSP}. The procedure is summarized in Algorithm~\ref{alg:btspp}.

\begin{algorithm}[t]
\caption{A $\lrA{2k+1}$-approximation for the \ac{BTSPP}}
\label{alg:btspp}
\begin{algorithmic}[1]
\REQUIRE The \ac{BTSPP} instance $\G=\lrA{\V,\E}$ with two terminal vertices $s$ and $t$.
\ENSURE A Hamiltonian path between $s$ and $t$ in $\G$.

\STATE Initialize $\O\coloneqq\emptyset$.\label{alg1s1}

\FOR{each irreducible state sequence $\zeta$ of length at most $f\lrA{k}$ and its corresponding path $\P_\zeta$ in $\G$}\label{alg1s2}
\STATE Let $\P \coloneqq \P_\zeta$.\label{alg1s3}
    \FOR{each $i \in\lrB{k}$ with $\overline{\V}_i\neq \emptyset$ ($\overline{\V}_i \coloneqq \V_i \setminus \V\lra{\P_\zeta}$)}\label{alg1s4}
            \STATE Identify a non-trivial state $\lra{\V_i, u_i, v_i}$ in $\zeta$.\label{alg1s5}

            \STATE Compute a Hamiltonian cycle $\CC_i$ in $\G\lrb{\overline{\V}_i \cup\lrc{u_i, v_i}}$ using the double-tree algorithm~\cite{WS:2011:book}, and then obtain a Hamiltonian path $\P_i$ between $u_i$ and $v_i$ in $\G\lrb{\overline{\V}_i \cup\lrc{u_i, v_i}}$ by shortcutting the walk $\lrd{u_i,\ldots,v_i}$ formed by the edges in $\E\lra{\CC_i}\uplus\lrc{u_iv_i}$.\label{alg1s6}

            \STATE Replace the edge $u_iv_i$ in $\P$ with the path $\P_i$.\label{alg1s7}
    \ENDFOR\label{alg1s8}
    \STATE $\O\coloneqq\O\cup\lrC{\P}$.\label{alg1s9}
\ENDFOR\label{alg1s10}
\RETURN $\arg\min_{\P \in \O} \c\lrA{\P}$.\label{alg1s11}
\end{algorithmic}
\end{algorithm}

\begin{lemma}\label{lem:btspp_approx}
For the \ac{BTSPP}, Algorithm~\ref{alg:btspp} is a $\lrA{2k+1}$-approximation algorithm with a running time of $N^{O\lra{2^kk!}}$.
\end{lemma}
\begin{IEEEproof}
Recall that $\pi^*$ denotes an optimal solution to the \ac{BTSPP}. 
By Lemma~\ref{lem:bound}, $\pi^*$ can be legally compressed into an irreducible state sequence $\zeta^*$ of length at most $f\lrA{k}$. 
By line~\ref{alg1s2}, Algorithm~\ref{alg:btspp} enumerates all possible paths corresponding to such valid state sequences. Hence, it will eventually consider the corresponding path $\P_{\zeta^*}$. 
Since $\P_{\zeta^*}$ is obtained via legal compressions (shortcutting internal vertices within the same region), the biased triangle inequality guarantees that 
\begin{equation}\label{alg1eq1}
\c\lra{\P_{\zeta^*}} \leq \c\lra{\pi^*}.
\end{equation}

For each $\V_i$ with $\overline{\V}_i \neq \emptyset$, by lines~\ref{alg1s4}--\ref{alg1s8}, the algorithm computes a Hamiltonian cycle $\CC_i$ in $\G\lrb{\overline{\V}_i \cup\lrc{u_i, v_i}}$. 
Since $\G\lrB{\V_i}$ is metric under $\c$, the double-tree algorithm ensures that
\begin{equation}\label{alg1eq2}
\c\lra{\CC_i} \leq 2 \c\lra{\T^*_{i}},
\end{equation}
where $\T^*_{i}$ is a minimum-cost spanning tree in $\G\lrb{\overline{\V}_i \cup\lrc{u_i, v_i}}$.

Recall that $\pi^*_{\V_i}$ denotes a minimum-cost Hamiltonian path in $\G\lrB{\V_i}$.
By Lemma~\ref{bound} and the triangle inequality, we have
\begin{equation}\label{alg1eq3}
\c\lra{\T^*_i} \leq \c\lra{\pi^*_{\V_i}} \leq \c\lra{\pi^*}.
\end{equation}

By line~\ref{alg1s6} and the triangle inequality, the Hamiltonian path $\P_i$ extracted between $u_i$ and $v_i$ satisfies
\begin{equation}\label{alg1eq4}
\c\lra{\P_i} \leq \c\lra{\CC_i} + \c\lra{u_i, v_i}.
\end{equation}

By line~\ref{alg1s7}, the patching operation replaces the edge $u_iv_i$ with $\P_i$. 
Therefore, the cost of the final path $\P$ is at most
\begin{align}
\c\lrA{\P} &\leq \c\lra{\P_{\zeta^*}} - \sum_{i: \overline{\V}_i \neq \emptyset} \c\lra{u_i, v_i} + \sum_{i: \overline{\V}_i \neq \emptyset} \c\lra{\P_i} \nonumber\\
&\leq \c\lra{\P_{\zeta^*}} - \sum_{i: \overline{\V}_i \neq \emptyset} \c\lra{u_i, v_i} + \sum_{i: \overline{\V}_i \neq \emptyset}\lra{\c\lra{\CC_i} + \c\lra{u_i, v_i}} \nonumber \\
&\leq \c\lra{\P_{\zeta^*}} + \sum_{i: \overline{\V}_i \neq \emptyset} \c\lra{\CC_i}.\label{alg1eq5}
\end{align}

Combining this with \eqref{alg1eq1}--\eqref{alg1eq5}, we obtain
\begin{equation}\label{alg1eq6}
\c\lrA{\P} \leq \c\lra{\pi^*} + \sum_{i=1}^k 2\c\lra{\pi^*} =\lrA{2k + 1} \c\lra{\pi^*}.
\end{equation}
Thus, the algorithm achieves an approximation ratio of $2k+1$.

Regarding the running time, enumerating all paths whose state sequences have length at most $f\lrA{k}=O\lra{2^kk!}$ takes $N^{O\lra{2^kk!}}$ time. For each enumeration, identifying unvisited vertices and running the double-tree algorithm takes $O\lra{N^2}$ time. Therefore, the total running time is bounded by $N^{O\lra{2^kk!}}$, which is polynomial for any constant $k$.
\end{IEEEproof}

\subsection{An $\lrA{\alpha k+1}$-Approximation Algorithm for the \ac{BTSP}}

For the \ac{BTSP}, our goal is to find a minimum-cost Hamiltonian cycle in $\G$. 
We instantiate the general framework by utilizing any $\alpha$-approximation algorithm for the metric \ac{TSP}.
The procedure is summarized in Algorithm~\ref{alg:btsp}.

\begin{algorithm}[t]
\caption{An $\lrA{\alpha k+1}$-approximation for the \ac{BTSP}}
\label{alg:btsp}
\begin{algorithmic}[1]
\REQUIRE The \ac{BTSP} instance $\G=\lrA{\V,\E}$.
\ENSURE A Hamiltonian cycle in $\G$.

\STATE Initialize $\O\coloneqq\emptyset$.\label{alg2s1}

\FOR{each irreducible state sequence $\zeta$ of length at most $f\lrA{k}$ and its corresponding path $\P_\zeta$ in $\G$}\label{alg2s2}
    \STATE Let $s$ and $t$ be the terminals of $\P_\zeta$. Form a cycle $\CC_\zeta$ by adding the edge $st$ to $\P_\zeta$, and let $\CC \coloneqq \CC_\zeta$.\label{alg2s3}
    \FOR{each $i \in\lrB{k}$ with $\overline{\V}_i\neq \emptyset$ ($\overline{\V}_i \coloneqq \V_i \setminus \V\lra{\CC_\zeta}$)}\label{alg2s4}
            \STATE Identify a non-trivial state $\lra{\V_i, u_i, v_i}$ in $\zeta$.\label{alg2s5}

            \STATE Compute a Hamiltonian cycle $\CC_i$ in $\G\lrb{\overline{\V}_i \cup\lrc{u_i, v_i}}$ using an $\alpha$-approximation algorithm, and then obtain a Hamiltonian path $\P_i$ between $u_i$ and $v_i$ in $\G\lrb{\overline{\V}_i \cup\lrc{u_i, v_i}}$ by shortcutting the walk $\lrd{u_i,\ldots,v_i}$ formed by the edges in $\E\lra{\CC_i}\uplus\lrc{u_iv_i}$.\label{alg2s6}

            \STATE Replace the edge $u_iv_i$ in $\CC$ with the path $\P_i$.\label{alg2s7}
    \ENDFOR\label{alg2s8}
    \STATE $\O\coloneqq\O\cup\lrc{\CC}$.\label{alg2s9}
\ENDFOR\label{alg2s10}
\RETURN $\arg\min_{\CC \in \O} \c\lrA{\CC}$.\label{alg2s11}
\end{algorithmic}
\end{algorithm}

\begin{lemma}\label{lem:btsp_approx}
For the \ac{BTSP}, Algorithm~\ref{alg:btsp} is an $\lrA{\alpha k+1}$-approximation algorithm with a running time of $N^{O\lra{2^kk!}}$.
\end{lemma}
\begin{IEEEproof}
Recall that $\sigma^*$ denotes an optimal solution to the \ac{BTSP}. 
Since state sequences are defined for paths, we can delete an arbitrary edge $st$ from $\sigma^*$ to yield a Hamiltonian path $\pi$. 
By Lemma~\ref{lem:bound}, $\pi$ can be legally compressed into an irreducible state sequence $\zeta^*$ of length at most $f\lrA{k}$. Let $\P_{\zeta^*}$ be the path corresponding to $\zeta^*$. 
Because legal compressions only shortcut internal vertices, they do not alter the terminals of the path. 
Consequently, the terminals of $\P_{\zeta^*}$ remain exactly $s$ and $t$.
Reconnecting the vertices $s$ and $t$ of $\P_{\zeta^*}$ with the edge $st$ forms a cycle $\CC_{\zeta^*}$. 
By lines~\ref{alg2s2}--\ref{alg2s3}, Algorithm~\ref{alg:btsp} enumerates all possible cycles constructed from such valid state sequences.
Hence, it will eventually consider $\CC_{\zeta^*}$. 
Since $\P_{\zeta^*}$ is obtained via legal compressions, the biased triangle inequality guarantees that $\c\lra{\P_{\zeta^*}} \leq \c\lrA{\pi}$. Therefore, the cost of the cycle $\CC_{\zeta^*}$ satisfies
\begin{equation}\label{alg2eq1}
\c\lra{\CC_{\zeta^*}} = \c\lra{\P_{\zeta^*}} + \c\lrA{s,t} \leq \c\lrA{\pi} + \c\lrA{s,t} = \c\lra{\sigma^*}.
\end{equation}

For each $\V_i$ with $\overline{\V}_i \neq \emptyset$, by lines~\ref{alg2s4}--\ref{alg2s8}, the algorithm computes a Hamiltonian cycle $\CC_i$ in $\G\lrb{\overline{\V}_i \cup\lrc{u_i, v_i}}$. 
Since $\G\lrB{\V_i}$ is metric under $\c$, the $\alpha$-approximation algorithm ensures that
\begin{equation}\label{alg2eq2}
\c\lra{\CC_i} \leq \alpha \c\lra{\sigma^*_{i}},
\end{equation}
where $\sigma^*_{i}$ is an optimal Hamiltonian cycle in $\G\lrb{\overline{\V}_i \cup\lrc{u_i, v_i}}$.

Recall that $\sigma^*_{\V_i}$ denotes a minimum-cost Hamiltonian cycle in $\G\lrB{\V_i}$.
By Lemma~\ref{bound} and the triangle inequality, we have
\begin{equation}\label{alg2eq3}
\c\lra{\sigma^*_{i}} \leq \c\lra{\sigma^*_{\V_i}} \leq \c\lra{\sigma^*}.
\end{equation}

By line~\ref{alg2s6} and the triangle inequality, the Hamiltonian path $\P_i$ extracted between $u_i$ and $v_i$ satisfies
\begin{equation}\label{alg2eq4}
\c\lra{\P_i} \leq \c\lra{\CC_i} + \c\lra{u_i, v_i}.
\end{equation}

By line~\ref{alg2s7}, the patching operation replaces the edge $u_iv_i$ with $\P_i$. 
Therefore, the cost of the final cycle $\CC$ is at most
\begin{align}
\c\lrA{\CC} &\leq \c\lra{\CC_{\zeta^*}} - \sum_{i: \overline{\V}_i \neq \emptyset} \c\lra{u_i, v_i} + \sum_{i: \overline{\V}_i \neq \emptyset} \c\lra{\P_i} \nonumber\\
&\leq \c\lra{\CC_{\zeta^*}} - \sum_{i: \overline{\V}_i \neq \emptyset} \c\lra{u_i, v_i} + \sum_{i: \overline{\V}_i \neq \emptyset}\lra{\c\lra{\CC_i} + \c\lra{u_i, v_i}} \nonumber \\
&\leq \c\lra{\CC_{\zeta^*}} + \sum_{i: \overline{\V}_i \neq \emptyset} \c\lra{\CC_i}.\label{alg2eq5}
\end{align}

Combining this with \eqref{alg2eq1}--\eqref{alg2eq5}, we obtain
\begin{equation}\label{alg2eq6}
\c\lrA{\CC} \leq \c\lra{\sigma^*} + \sum_{i=1}^k \alpha\c\lra{\sigma^*} =\lrA{\alpha k + 1} \c\lra{\sigma^*}.
\end{equation}
Thus, the algorithm achieves an approximation ratio of $\alpha k+1$.

Regarding the running time, enumerating all paths corresponding to valid irreducible state sequences of length at most $f\lrA{k}=O\lra{2^kk!}$ and forming their respective initial cycles takes $N^{O\lra{2^kk!}}$ time. For each enumeration, identifying unvisited vertices and running the $\alpha$-approximation algorithm takes polynomial time $O\lrA{1}$. Therefore, the total running time remains bounded by $N^{O\lra{2^kk!}}$, which is polynomial for any constant $k$.
\end{IEEEproof}

\section{The Algorithm for the \ac{BTSPP} with $k=2$}\label{sec:4}
In this section, we study the \ac{BTSPP} with $k=2$ and propose an $\lrA{\alpha+2+\varepsilon}$-approximation algorithm for the \ac{BTSPP} with polynomial time, where $\varepsilon>0$ denotes any positive constant.
The algorithm relies on refined state compression techniques and requires carefully reconstructing Hamiltonian paths from compressed state sequences.
We first present a reconstruction procedure that will be used throughout the algorithm.

\begin{lemma}\label{path}
Consider a metric graph $\G=\lrA{\V,\E}$ with cost function $c$.
Given any distinct vertices $x,y\in \V$ and any $\V'\subseteq \V$ with $\V'\supseteq\lrc{x,y}$, there exists an $O\lra{\size{\V}^2}$-time algorithm that computes a Hamiltonian path $\pi$ between $x$ and $y$ in $\G\lrB{\V'}$ such that $c\lrA{\pi}\leq 2c\lra{\pi_\G^*}$, where $\pi_\G^*$ is a minimum-cost Hamiltonian path in $\G$.
\end{lemma}

\begin{IEEEproof}
We compute a minimum-cost spanning tree $\T$ in $\G\lrb{\V'}$.
Since $\T$ is a tree, there is a unique path $\P$ between $x$ and $y$ in $\T$.
By doubling all edges in $\E\lrA{\T}\setminus \E\lrA{\P}$, we obtain a multigraph in which every vertex has even degree except $x$ and $y$.
Therefore, a walk between $x$ and $y$ that traverses every edge in the multigraph exactly once can be found in linear time~\cite{CLRS:2022:book}.
By shortcutting repeated vertices in this walk, we obtain a Hamiltonian path $\pi$ between $x$ and $y$ in $\G\lrb{\V'}$.
The triangle inequality guarantees that the shortcutting operation does not increase the cost.
Hence, $c\lrA{\pi}\leq 2c\lrA{\T}$.
Since $\pi_\G^*$ can be shortcut to obtain a Hamiltonian path $\hat{\pi}$ in $\G\lrb{\V'}$ with $c\lra{\hat{\pi}}\leq c\lra{\pi_\G^*}$, and $\hat{\pi}$ is also a spanning tree of $\G\lrb{\V'}$, the minimum spanning tree $\T$ satisfies $c\lrA{\T} \leq c\lra{\hat{\pi}}\leq c\lra{\pi_\G^*}$.
Therefore, $c\lrA{\pi} \leq 2c\lra{\pi_\G^*}$.

The running time is dominated by computing the minimum spanning tree, which takes $O\lra{\size{\V}^2}$ time~\cite{CLRS:2022:book}.
\end{IEEEproof}

\subsection{The Algorithm}

Fix a large threshold $L$ (defined later) such that $L$ is even and $L=O\lrA{1/\varepsilon}$.
For convenience, we may assume that 
\begin{equation}\label{ass}
\size{\V}\geq 2L,
\end{equation}
as otherwise, the problem can be solved by brute-force enumeration of all Hamiltonian paths in constant time $O\lra{\size{\V}!}=O\lra{L!}$.

We call edges $xy\in \E$ with $x\in \V_1$ and $y\in \V_2$ \emph{cross edges}. 
We then fix a minimum-cost Hamiltonian path $\pi^*=\lrd{s,\ldots, t}$ in $\G$, i.e., an optimal solution to the \ac{BTSPP}, and assume without loss of generality that $s\in \V_1$ and $t\in \V_\theta$, where $\theta\in\lrc{1,2}$. Note that $\V_\theta=\V_1$ if $s$ and $t$ are in the same subset, and $\V_\theta=\V_2$ otherwise.

For any state subsequences $\zeta$ and $\zeta'$, we use $\zeta\circ\zeta'$ to denote their concatenation, where $\zeta$ appears before $\zeta'$.

Let $\zeta_{\pi^*}=\lrd{\lra{\V_1,s,s},\ldots,\lra{\V_\theta,t,t}}$ denote the initial state sequence of $\pi^*$.
Clearly, $\zeta_{\pi^*}$ has length at least $2L$ by \eqref{ass}.
Let $\zeta_1$ be the subsequence of $\zeta_{\pi^*}$ consisting of the first $L$ states, and $\zeta_2$ be the subsequence of $\zeta_{\pi^*}$ consisting of the remaining states. Hence, $\zeta_{\pi^*} = \zeta_1 \circ \zeta_2$.

\paragraph{Case~1: $\zeta_1$ contains two consecutive states associated with $\V_{3-\theta}$}
In this case, $\zeta_1$ may contain multiple consecutive states associated with $\V_{3-\theta}$ of length at least 2, separated by states associated with $\V_\theta$. 
We consider the last maximal consecutive states associated with $\V_{3-\theta}$ of length at least 2, denoted by
\begin{equation}\label{c1e1}
\zeta'_1 =\lrd{\lra{\V_{3-\theta}, u, u},\ldots,\lra{\V_{3-\theta}, x, x}},
\end{equation}
where $u,x\in \V_{3-\theta}$ and $u\neq x$.

Let $\zeta'_2$ denote the subsequence of $\zeta_{\pi^*}$, starting from the first state associated with $\V_\theta$ after the state $\lra{\V_{3-\theta}, x, x}$, denoted by $\lra{\V_\theta, z, z}$, and ending at the last state $\lra{\V_\theta, t, t}$.
Note that the subsequence of $\zeta_{\pi^*}$ strictly between $\zeta'_1$ and $\zeta'_2$, denoted by $\zeta''_1$, contains only states associated with $\V_{3-\theta}$. 
Let the last state of $\zeta''_1$ be $\lra{\V_{3-\theta}, y, y}$.
Then,
\begin{equation}\label{c1e2}
\zeta'_1\circ\zeta''_1 =\lrd{\lra{\V_{3-\theta}, u, u},\ldots,\lra{\V_{3-\theta}, y, y}},
\end{equation}
\begin{equation}\label{c1e3}
\zeta'_2 =\lrd{\lra{\V_\theta, z, z},\ldots,\lra{\V_\theta, t, t}},
\end{equation}
and
\begin{equation}\label{c1e3+}
\zeta'_1\circ\zeta''_1\circ\zeta'_2 = \zeta_{\pi^*}.
\end{equation}

Note that if $\zeta''_1$ is empty, then define $\zeta'_1\circ\zeta''_1 = \zeta'_1$ and $y=x$.
Moreover, we may have $z=t$.
We can legally compress $\zeta_{\pi^*}$ by applying \ac{MO}-2 to the subsequences $\zeta'_1\circ\zeta''_1$ and $\zeta'_2$ (if $z\neq t$), obtaining a state sequence, denoted by $\zeta'_{\pi^*}$ such that
\begin{equation}\label{c1e4}
\zeta'_{\pi^*} =\lrd{\ldots,\lra{\V_{3-\theta}, u, y},\lra{\V_\theta, z, t}}.
\end{equation}

\begin{lemma}\label{lm1}
For the \ac{BTSPP} with $k=2$ in Case~1, there exists a polynomial-time $\lrA{\alpha+2+\varepsilon}$-approximation algorithm.
\end{lemma}
\begin{IEEEproof}
The algorithm begins by enumerating $\zeta'_{\pi^*}$. 
Since $\zeta'_{\pi^*}$ is obtained via legal compression of $\zeta_{\pi^*}$, by \eqref{c1e2}--\eqref{c1e4}, the corresponding path $\P_{\zeta'_{\pi^*}}$ in $\G$ satisfies
\begin{equation}\label{lm1eq1}
\begin{split}
\c\lra{\P_{\zeta'_{\pi^*}}} &= \c\lra{\P_{\zeta_{\pi^*}}} - \c\lra{\P_{\zeta_{\pi^*}}\lrA{u,y}} - \c\lra{\P_{\zeta_{\pi^*}}\lrA{z,t}}\\
&\quad\quad + \c\lrA{u,y}+\c\lrA{z,t}\\
&\leq \c\lra{\pi^*}-\c\lra{\P_{\zeta_{\pi^*}}\lrA{z,t}}+\c\lrA{z,t},
\end{split}
\end{equation}
where the equality follows from \eqref{c1e3+} and the inequality follows from $\c\lrA{u,y}\leq \c\lra{\P_{\zeta_{\pi^*}}\lrA{u,y}}$ by the triangle inequality and the fact that $\c\lra{\P_{\zeta_{\pi^*}}}=\c\lra{\pi^*}$.

Note that $\P_{\zeta'_{\pi^*}}$ is not a Hamiltonian path in $\G$, since it does not visit the internal vertices in the paths $\P_{\zeta_{\pi^*}}\lrA{u,y}$ and $\P_{\zeta_{\pi^*}}\lrA{z,t}$ (if $z\neq t$) by definition.
Denote the set of these vertices in $\V_{3-\theta}$ and $\V_\theta$ by $\V'_{3-\theta}$ and $\V'_\theta$, respectively.

To cover the vertices in $\V'_{3-\theta}$, we replace the edge $uy$ in $\P_{\zeta'_{\pi^*}}$ with the path $\P\lrA{u,y}$ in $\G\lrB{\V'_{3-\theta}\cup\lrc{u,y}}$ obtained by applying the algorithm in Lemma~\ref{path} with the metric graph $\G\lrB{\V_{3-\theta}}$ and the vertex set $\V'_{3-\theta}\cup\lrc{u,y}$ and the vertices $u$ and $y$.
By Lemma~\ref{path}, we have
\begin{equation}\label{lm1eq2}
\c\lra{\P\lrA{u,y}} \leq 2\c\lra{\pi^*_{\V_{3-\theta}}}\leq 2 \c\lra{\pi^*},
\end{equation}
where the first inequality follows from the fact that $\pi^*_{\V_{3-\theta}}$ is a Hamiltonian path in $\G\lrb{\V_{3-\theta}}$ and the second from Lemma~\ref{bound}.

Moreover, if $z\neq t$, to cover the vertices in $\V'_{\theta}$, we further replace the edge $zt$ in $\P_{\zeta'_{\pi^*}}$ with an $\lrA{\alpha+\varepsilon}$-approximate Hamiltonian path between $z$ and $t$ in $\G\lrb{\V'_\theta\cup\lrc{z,t}}$ by applying the algorithm in~\cite{TVZ:2022:article}.

By \eqref{c1e3} and \eqref{c1e4}, a valid Hamiltonian path between $z$ and $t$ in $\G\lrb{\V'_\theta\cup\lrc{z,t}}$ can be obtained via shortcutting $\P_{\zeta_{\pi^*}}\lrA{z,t}$ without increasing the cost. 
Thus, we have
\begin{equation}\label{lm1eq3}
\c\lra{\P\lrA{z,t}} \leq\lrA{\alpha+\varepsilon} \c\lra{\P_{\zeta_{\pi^*}}\lrA{z,t}}.
\end{equation}

Denote the resulting path by $\pi$, which is a Hamiltonian path in $\G$ by definition.
By \eqref{lm1eq1}--\eqref{lm1eq3}, we have 
\begin{equation}\label{lm1eq4}
\begin{split}
\c\lrA{\pi} &= \c\lra{\P_{\zeta'_{\pi^*}}} - \c\lrA{u,y} - \c\lrA{z,t}\\
&\quad\quad+ \c\lra{\P\lrA{u,y}}+ \c\lra{\P\lrA{z,t}}\\
&\leq \c\lra{\P_{\zeta'_{\pi^*}}}  - \c\lrA{z,t} +\c\lra{\P\lrA{u,y}}+\c\lra{\P\lrA{z,t}}\\
&\leq \c\lra{\pi^*}-\c\lra{\P_{\zeta_{\pi^*}}\lrA{z,t}}+ \c\lra{\P\lrA{u,y}}+\c\lra{\P\lrA{z,t}}\\
&\leq 3\c\lra{\pi^*}+\lrA{\alpha-1+\varepsilon}\c\lra{\P_{\zeta_{\pi^*}}\lrA{z,t}}\\
&\leq\lrA{\alpha+2+\varepsilon}\c\lra{\pi^*},
\end{split}
\end{equation}
where the second inequality follows from \eqref{lm1eq1}, the third from \eqref{lm1eq2} and \eqref{lm1eq3}, and the last from  $\alpha+\varepsilon \geq 1$ and $\c\lra{\P_{\zeta_{\pi^*}}\lrA{z,t}}\leq \c\lra{\P_{\zeta_{\pi^*}}} =\c\lra{\pi^*}$ by the triangle inequality.

Therefore, the above algorithm produces an $\lrA{\alpha+2+\varepsilon}$-approximate solution for the \ac{BTSPP} with $k=2$.

For the running time, it is dominated by enumerating $\zeta'_{\pi^*}$.
By \eqref{c1e1} and \eqref{c1e4}, we have $\size{\zeta'_{\pi^*}}\leq \size{\zeta'_1} = L$.
Therefore, the overall running time is $N^{O\lrA{L}}$.
\end{IEEEproof}

\paragraph{Case~2: $\zeta_1$ contains no two consecutive states associated with $\V_{3-\theta}$ but contains two consecutive states associated with $\V_\theta$}
In this case, $\zeta_1$ may contain multiple consecutive states associated with $\V_\theta$ of length at least 2, separated by states associated with $\V_{3-\theta}$. 
We consider the last maximal consecutive states associated with $\V_\theta$ of length at least 2, denoted by 
\[
\zeta'_1 =\lrd{\lra{\V_\theta, u, u},\ldots,\lra{\V_\theta, x, x}},
\]
where $u,x\in \V_\theta$ and $u\neq x$.

\paragraph{Subcase~2.1: All states are associated with $\V_\theta$ after $\lra{\V_\theta, x, x}$ in $\zeta_{\pi^*}$}
In this case, $\lra{\V_\theta, x, x}$ must be the last state of $\zeta_1$, and all states in $\zeta_2$ are associated with $\V_\theta$.

We can legally compress $\zeta_{\pi^*}$ by applying \ac{MO}-1 to the subsequence $\zeta'_1\circ\zeta_2$, obtaining a state sequence, denoted by $\zeta'_{\pi^*}$ such that
\[
\zeta'_{\pi^*} =\lrd{\ldots,\lra{\V_\theta, u, t}}.
\]

\begin{lemma}\label{lm2}
For the \ac{BTSPP} with $k=2$ in Subcase~2.1, there exists a polynomial-time $\lrA{\alpha+\varepsilon}$-approximation algorithm.
\end{lemma}
\begin{IEEEproof}
The algorithm begins by enumerating $\zeta'_{\pi^*}$. 
Since $\zeta'_{\pi^*}$ is obtained via legal compression of $\zeta_{\pi^*}$ by applying \ac{MO}-1 to the subsequence $\zeta'_1\circ\zeta_2$, the corresponding path $\P_{\zeta'_{\pi^*}}$ in $\G$ satisfies
\begin{equation}\label{lm2eq1}
\begin{split}
\c\lra{\P_{\zeta'_{\pi^*}}} &= \c\lra{\P_{\zeta_{\pi^*}}} - \c\lra{\P_{\zeta_{\pi^*}}\lrA{u,t}} + \c\lrA{u,t} \\
&= \c\lra{\pi^*} - \c\lra{\P_{\zeta_{\pi^*}}\lrA{u,t}} + \c\lrA{u,t},
\end{split}
\end{equation}
where the equality follows from $\c\lra{\P_{\zeta_{\pi^*}}} = \c\lra{\pi^*}$.

Note that $\P_{\zeta'_{\pi^*}}$ is not a Hamiltonian path in $\G$, since it does not visit the internal vertices in the path $\P_{\zeta_{\pi^*}}\lrA{u,t}$ by definition. 
Denote the set of these internal vertices in $\V_{3-\theta}$ and $\V_\theta$ by $\V'_{3-\theta}$ and $\V'_\theta$, respectively.
In Subcase~2.1, all states appearing after $\lra{\V_\theta, x, x}$ in $\zeta_{\pi^*}$ (including $\zeta_2$) are associated with $\V_\theta$. This implies that the path $\P_{\zeta_{\pi^*}}\lrA{u,t}$ consists entirely of vertices in $\V_\theta$. Therefore, we have $\V'_{3-\theta}=\emptyset$.

To cover the unvisited vertices in $\V'_\theta$, we replace the edge $ut$ in $\P_{\zeta'_{\pi^*}}$ with an $\lrA{\alpha+\varepsilon}$-approximate path $\P\lrA{u,t}$ between $u$ and $t$ in the induced subgraph $\G\lrb{\V'_\theta\cup\lrc{u,t}}$. 
Since $\P_{\zeta_{\pi^*}}\lrA{u,t}$ is a valid path between $u$ and $t$ that exactly covers all vertices in $\V'_\theta\cup\lrc{u,t}$, the cost of the path $\P\lrA{u,t}$ is at most
\begin{equation}\label{lm2eq2}
\c\lra{\P\lrA{u,t}} \leq\lrA{\alpha+\varepsilon} \c\lra{\P_{\zeta_{\pi^*}}\lrA{u,t}}.
\end{equation}

By replacing the edge $ut$ with $\P\lrA{u,t}$ in $\P_{\zeta'_{\pi^*}}$, we obtain a valid Hamiltonian path $\pi$ in $\G$. The total cost of $\pi$ can be bounded as follows:
\begin{equation}\label{lm2eq3}
\begin{split}
\c\lrA{\pi} &= \c\lra{\P_{\zeta'_{\pi^*}}} - \c\lrA{u,t} + \c\lra{\P\lrA{u,t}} \\
&= \c\lra{\pi^*} - \c\lra{\P_{\zeta_{\pi^*}}\lrA{u,t}} + \c\lra{\P\lrA{u,t}} \\
&\leq \c\lra{\pi^*} - \c\lra{\P_{\zeta_{\pi^*}}\lrA{u,t}} +\lrA{\alpha+\varepsilon} \c\lra{\P_{\zeta_{\pi^*}}\lrA{u,t}} \\
&= \c\lra{\pi^*} +\lrA{\alpha+\varepsilon - 1} \c\lra{\P_{\zeta_{\pi^*}}\lrA{u,t}} \\
&\leq \c\lra{\pi^*} +\lrA{\alpha+\varepsilon - 1} \c\lra{\pi^*} \\
&=\lrA{\alpha+\varepsilon} \c\lra{\pi^*},
\end{split}
\end{equation}
where the second equality follows from \eqref{lm2eq1}, the first inequality from \eqref{lm2eq2}, and the second inequality from $\alpha+\varepsilon \geq 1$ and $\c\lra{\P_{\zeta_{\pi^*}}\lrA{u,t}} \leq \c\lra{\P_{\zeta_{\pi^*}}} = \c\lra{\pi^*}$ by the triangle inequality.

Similarly, the running time of the algorithm is dominated by enumerating $\zeta'_{\pi^*}$ whose length is bounded by a constant $O\lrA{L}$. 
Therefore, the overall running time is $N^{O\lrA{L}}$.
\end{IEEEproof}

\paragraph{Subcase~2.2: There exists a state associated with $\V_{3-\theta}$ after $\lra{\V_\theta, x, x}$ in $\zeta_{\pi^*}$}
Let $\zeta'_2$ denote the subsequence of $\zeta_{\pi^*}$, starting from the first state associated with $\V_{3-\theta}$ after the state $\lra{\V_\theta, x, x}$, denoted by $\lra{\V_{3-\theta}, v, v}$, and ending at the last state associated with $\V_{3-\theta}$, denoted by $\lra{\V_{3-\theta}, v', v'}$.

Note that the subsequence of $\zeta_{\pi^*}$ strictly between $\zeta'_1$ and $\zeta'_2$, denoted by $\zeta''_1$, contains only states associated with $\V_\theta$, and the subsequence of $\zeta_{\pi^*}$ strictly after $\zeta'_2$, denoted by $\zeta''_2$, contains only states associated with $\V_\theta$.
Let the last state of $\zeta''_1$ be $\lra{\V_\theta, y, y}$ and the first state of $\zeta''_2$ be $\lra{\V_\theta, z, z}$.
Then,
\[
\zeta'_1\circ\zeta''_1 =\lrd{\lra{\V_\theta, u, u},\ldots,\lra{\V_\theta, y, y}},
\]
\[
\zeta'_2 =\lrd{\lra{\V_{3-\theta}, v, v},\ldots,\lra{\V_{3-\theta}, v', v'}},
\]
and
\[
\zeta''_2 =\lrd{\lra{\V_\theta, z, z},\ldots,\lra{\V_\theta, t, t}}.
\]

Note that if $\zeta''_1$ is empty, then define $\zeta'_1\circ\zeta''_1 = \zeta'_1$ and $y=x$.
Moreover, we may have $v=v'$ or $z=t$.
We can legally compress $\zeta_{\pi^*}$ by applying \ac{MO}-2 to the subsequences $\zeta'_1\circ\zeta''_1$, $\zeta'_2$ (if $v\neq v'$), and $\zeta''_2$ (if $z\neq t$), obtaining a state sequence, denoted by $\zeta'_{\pi^*}$ such that
\[
\zeta'_{\pi^*} =\lrd{\ldots,\lra{\V_\theta, u, y},\lra{\V_{3-\theta}, v, v'},\lra{\V_\theta, z, t}}.
\]

\begin{lemma}\label{lm3}
For the \ac{BTSPP} with $k=2$ in Subcase~2.2, there exists a polynomial-time $\lrA{\alpha+2+\varepsilon}$-approximation algorithm.
\end{lemma}
\begin{IEEEproof}
The algorithm begins by enumerating $\zeta'_{\pi^*}$. 
Since $\zeta'_{\pi^*}$ is obtained via legal compression of $\zeta_{\pi^*}$ by applying \ac{MO}-2 to the subsequences $\zeta'_1\circ\zeta''_1$, $\zeta'_2$, and $\zeta''_2$, the corresponding path $\P_{\zeta'_{\pi^*}}$ in $\G$ satisfies
\begin{equation}\label{lm3eq1}
\begin{split}
\c\lra{\P_{\zeta'_{\pi^*}}} &= \c\lra{\P_{\zeta_{\pi^*}}} - \c\lra{\P_{\zeta_{\pi^*}}\lrA{u,y}} - \c\lra{\P_{\zeta_{\pi^*}}\lra{v,v'}}\\
&\quad\quad - \c\lra{\P_{\zeta_{\pi^*}}\lrA{z,t}}  + \c\lrA{u,y} + \c\lra{v,v'} + \c\lrA{z,t}\\
&\leq \c\lra{\pi^*} - \c\lra{\P_{\zeta_{\pi^*}}\lra{v,v'}} + \c\lra{v,v'},
\end{split}
\end{equation}
where the inequality follows from $\c\lrA{z,t}\leq \c\lra{\P_{\zeta_{\pi^*}}\lrA{z,t}}$ and $\c\lrA{u,y}\leq \c\lra{\P_{\zeta_{\pi^*}}\lrA{u,y}}$ by the triangle inequality and the fact that $\c\lra{\P_{\zeta_{\pi^*}}} = \c\lra{\pi^*}$.

Note that $\P_{\zeta'_{\pi^*}}$ does not visit the internal vertices in the paths $\P_{\zeta_{\pi^*}}\lrA{u,y}$, $\P_{\zeta_{\pi^*}}\lra{v,v'}$, and $\P_{\zeta_{\pi^*}}\lrA{z,t}$. 
Denote the set of these vertices in $\V_{3-\theta}$ and $\V_\theta$ by $\V'_{3-\theta}$ and $\V'_\theta$, respectively.

To cover the unvisited vertices in $\V'_{3-\theta}$, we replace the edge $vv'$ in $\P_{\zeta'_{\pi^*}}$ with an $\lrA{\alpha+\varepsilon}$-approximate path $\P\lra{v,v'}$ between $v$ and $v'$ in $\G\lrb{\V'_{3-\theta}\cup\lrc{v,v'}}$.

Note that a valid Hamiltonian path between $v$ and $v'$ in $\G\lrb{\V'_{3-\theta}\cup\lrc{v,v'}}$ can be obtained via shortcutting $\P_{\zeta_{\pi^*}}\lra{v,v'}$ without increasing the cost.
Thus, we have
\begin{equation}\label{lm3eq2}
\c\lra{\P\lra{v,v'}} \leq\lrA{\alpha+\varepsilon}\c\lra{\P_{\zeta_{\pi^*}}\lra{v,v'}}.
\end{equation}

To cover the unvisited vertices in $\V'_\theta$, we replace the edge $uy$ in $\P_{\zeta'_{\pi^*}}$ with a path $\P\lrA{u,y}$ between $u$ and $y$ that covers exactly all vertices in $\V'_\theta\cup\lrc{u,y}$ using the algorithm in Lemma~\ref{path}. 
By Lemma~\ref{path}, we have
\begin{equation}\label{lm3eq3}
\c\lra{\P\lrA{u,y}}\leq 2\c\lra{\pi^*_{\V_{\theta}}}\leq 2\c\lra{\pi^*}.
\end{equation}

By replacing the edge $vv'$ with $\P\lra{v,v'}$ and the edge $uy$ with $\P\lrA{u,y}$ in $\P_{\zeta'_{\pi^*}}$, we obtain a valid Hamiltonian path $\pi$ in $\G$. The total cost of $\pi$ can be bounded as follows:
\begin{equation}\label{lm3eq4}
\begin{split}
\c\lrA{\pi} &= \c\lra{\P_{\zeta'_{\pi^*}}} - \c\lra{v,v'}  - \c\lrA{u,y} \\
&\quad\quad + \c\lra{\P\lra{v,v'}} + \c\lra{\P\lrA{u,y}}\\
&\leq \c\lra{\P_{\zeta'_{\pi^*}}} - \c\lra{v,v'}+ \c\lra{\P\lra{v,v'}} + \c\lra{\P\lrA{u,y}}\\
&\leq 3\c\lra{\pi^*} +\lrA{\alpha+\varepsilon - 1}\c\lra{\P_{\zeta_{\pi^*}}\lra{v,v'}} \\
&\leq\lrA{\alpha+2+\varepsilon} \c\lra{\pi^*},
\end{split}
\end{equation}
where the second inequality follows from \eqref{lm3eq1}, \eqref{lm3eq2}, and \eqref{lm3eq3}, and the last inequality from $\alpha+\varepsilon \geq 1$ and $\c\lra{\P_{\zeta_{\pi^*}}\lra{v,v'}} \leq \c\lra{\P_{\zeta_{\pi^*}}} = \c\lra{\pi^*}$ by the triangle inequality.

Similarly, the running time of the algorithm is dominated by enumerating $\zeta'_{\pi^*}$ whose length is bounded by $O\lrA{L}$. 
Furthermore, computing the $\lrA{\alpha+\varepsilon}$-approximate path $\P\lra{v,v'}$ and the path $\P\lrA{u,y}$ via Lemma~\ref{path} both run in polynomial time. Therefore, the overall running time is $N^{O\lrA{L}}$, which is polynomial. 
This completes the proof.
\end{IEEEproof}

\paragraph{Case~3: $\zeta_1$ contains no consecutive states associated with $\V_{3-\theta}$ or $\V_\theta$}
In this case, the states in $\zeta_1$ alternate between being associated with $\V_{3-\theta}$ and $\V_\theta$, starting with either one.
Specifically, since $L$ is even, if $\V_1= \V_\theta$, the states alternate between $\V_\theta$ and $\V_{3-\theta}$ exactly $L/2$ times; otherwise, the states alternate between $\V_{3-\theta}$ and $\V_\theta$ exactly $L/2$ times.
Therefore, in either case, there exists a subsequence of $\zeta_1$, denoted by $\zeta'_1$, such that the states in $\zeta'_1$ alternate between $\V_{3-\theta}$ and $\V_\theta$ exactly $L/2-1$ times, and the state $\lra{\V_1,s,s}$ is not contained in $\zeta'_1$.

Let
\[
\zeta'_1=\zeta^1_1\circ\zeta^2_1\circ\cdots\circ\zeta^{L/2-1}_1,
\]
where $\zeta^i_1=\lrd{\lra{\V_{3-\theta}, x_i, x_i},\lra{\V_\theta, y_i, y_i}}$.

The path $\P_{\zeta_{\pi^*}}$ visits the vertices $y_1, y_2, \ldots, y_{L/2-1}$ in that specific order. 
Since the subpaths $\P_{\zeta_{\pi^*}}\lra{y_{i-1}, y_i}$ for $i=2, \ldots, L/2-1$ are edge-disjoint segments of $\P_{\zeta_{\pi^*}}$, the sum of their costs is bounded by the total cost of $\pi^*$:
\[
\sum_{i=2}^{L/2-1} \c\lra{\P_{\zeta_{\pi^*}}\lra{y_{i-1}, y_i}} \leq \c\lra{\P_{\zeta_{\pi^*}}} = \c\lra{\pi^*}.
\]
Since there are $L/2-2$ such subpaths, by an averaging argument, there exists an $i^* \in\lrc{2, \ldots, L/2-1}$ such that
\[
\c\lra{\P_{\zeta_{\pi^*}}\lra{y_{i^*-1}, y_{i^*}}} \leq \frac{1}{L/2 - 2}\c\lra{\pi^*} = \frac{2}{L-4}\c\lra{\pi^*}.
\]
By the triangle inequality, we have
\begin{equation}\label{lm4eq_avg}
\c\lra{y_{i^*-1}, y_{i^*}} \leq \c\lra{\P_{\zeta_{\pi^*}}\lra{y_{i^*-1}, y_{i^*}}} \leq \frac{2}{L-4}\c\lra{\pi^*}.
\end{equation}

We legally compress $\zeta_{\pi^*}$ to obtain $\zeta'_{\pi^*}$ by applying \ac{MO}-1 to two subsequences.
First, we apply \ac{MO}-1 to the subsequence starting at $\lra{\V_{3-\theta}, x_{i^*-1}, x_{i^*-1}}$ and ending at $\lra{\V_{3-\theta}, x_{i^*}, x_{i^*}}$, which compresses it into the state $\lra{\V_{3-\theta}, x_{i^*-1}, x_{i^*}}$.
Second, we apply \ac{MO}-1 to the subsequence starting at $\lra{\V_\theta, y_{i^*}, y_{i^*}}$ and ending at $\lra{\V_\theta, t, t}$, compressing it into $\lra{\V_\theta, y_{i^*}, t}$.

\begin{lemma}\label{lm4}
For the \ac{BTSPP} with $k=2$ in Case~3, there exists a polynomial-time $\lra{\alpha+2+\frac{4}{L-4}+\varepsilon}$-approximation algorithm.
\end{lemma}
\begin{IEEEproof}
The algorithm enumerates $\zeta'_{\pi^*}$. 
By the definition of the compressions, the corresponding path $\P_{\zeta'_{\pi^*}}$ satisfies
\begin{equation}\label{lm4eq1}
\begin{split}
\c\lra{\P_{\zeta'_{\pi^*}}} &= \c\lra{\P_{\zeta_{\pi^*}}} - \c\lra{\P_{\zeta_{\pi^*}}\lra{x_{i^*-1}, x_{i^*}}} - \c\lra{\P_{\zeta_{\pi^*}}\lra{y_{i^*}, t}} \\
&\quad + \c\lra{x_{i^*-1}, x_{i^*}} + \c\lra{y_{i^*}, t} \\
&\leq \c\lra{\pi^*} - \c\lra{\P_{\zeta_{\pi^*}}\lra{y_{i^*}, t}} + \c\lra{y_{i^*}, t},
\end{split}
\end{equation}
where the inequality follows from $\c\lra{x_{i^*-1},x_{i^*}}\leq \c\lra{\P_{\zeta_{\pi^*}}\lra{x_{i^*-1}, x_{i^*}}}$ by the triangle inequality and the fact that $\c\lra{\P_{\zeta_{\pi^*}}} = \c\lra{\pi^*}$.

Note that $\P_{\zeta'_{\pi^*}}$ does not visit the internal vertices in the paths $\P_{\zeta_{\pi^*}}\lra{x_{i^*-1}, x_{i^*}}$ and $\P_{\zeta_{\pi^*}}\lra{y_{i^*}, t}$.
Denote the set of these unvisited vertices in $\V_{3-\theta}$ and $\V_\theta$ by $\V'_{3-\theta}$ and $\V'_\theta$, respectively.
Specifically, $y_{i^*-1} \in \V'_\theta$ because it is an internal vertex of $\P_{\zeta_{\pi^*}}\lra{x_{i^*-1}, x_{i^*}}$.

To cover the unvisited vertices in $\V'_{3-\theta}$, we replace the edge $x_{i^*-1}x_{i^*}$ in $\P_{\zeta'_{\pi^*}}$ with a path $\P\lra{x_{i^*-1}, x_{i^*}}$ between $x_{i^*-1}$ and $x_{i^*}$ that covers exactly all vertices in $\V'_{3-\theta}\cup\lrc{x_{i^*-1}, x_{i^*}}$. Using the algorithm in Lemma~\ref{path}, we have
\begin{equation}\label{lm4eq2}
\c\lra{\P\lra{x_{i^*-1}, x_{i^*}}} \leq 2\c\lra{\pi^*_{\V_{3-\theta}}} \leq 2\c\lra{\pi^*}.
\end{equation}

For the unvisited vertices in $\V'_\theta$, we first construct a path for the subset $\V'_\theta \setminus\lrc{y_{i^*-1}}$.
We obtain an $\lrA{\alpha+\varepsilon}$-approximate path $\P'\lra{y_{i^*}, t}$ between $y_{i^*}$ and $t$ in $\G\lrb{\V'_\theta \cup\lrc{y_{i^*}, t} \setminus\lrc{y_{i^*-1}}}$.
Since shortcutting $\P_{\zeta_{\pi^*}}\lra{y_{i^*}, t}$ yields a valid Hamiltonian path in this subgraph without increasing the cost, we have
\begin{equation}\label{lm4eq3}
\c\lra{\P'\lra{y_{i^*}, t}} \leq\lrA{\alpha+\varepsilon}\c\lra{\P_{\zeta_{\pi^*}}\lra{y_{i^*}, t}}.
\end{equation}

Then, we update $\P'\lra{y_{i^*}, t}$ to additionally cover $y_{i^*-1}$ by adding two copies of the edge $y_{i^*-1}y_{i^*}$ (i.e., traversing from $y_{i^*}$ to $y_{i^*-1}$ and back) and then shortcutting. 
Let this final path be $\P\lra{y_{i^*}, t}$, which now exactly covers $\V'_\theta\cup\lrc{y_{i^*}, t}$. The cost increases by at most $2\c\lra{y_{i^*-1}, y_{i^*}}$ by the triangle inequality. 
Thus, using \eqref{lm4eq_avg} and \eqref{lm4eq3}, we obtain
\begin{equation}\label{lm4eq4}
\begin{split}
\c\lra{\P\lra{y_{i^*}, t}} &\leq \c\lra{\P'\lra{y_{i^*}, t}} + 2\c\lra{y_{i^*-1}, y_{i^*}} \\
&\leq\lrA{\alpha+\varepsilon} \c\lra{\P_{\zeta_{\pi^*}}\lra{y_{i^*}, t}} + \frac{4}{L-4}\c\lra{\pi^*}.
\end{split}
\end{equation}

Finally, by replacing the edge $x_{i^*-1}x_{i^*}$ with $\P\lra{x_{i^*-1}, x_{i^*}}$ and the edge $y_{i^*}t$ with $\P\lra{y_{i^*}, t}$ in $\P_{\zeta'_{\pi^*}}$, we obtain a valid Hamiltonian path $\pi$ in $\G$.
We have
\begin{equation}\label{lm4eq5}
\begin{split}
\c\lrA{\pi} &= \c\lra{\P_{\zeta'_{\pi^*}}} - \c\lra{x_{i^*-1}, x_{i^*}} - \c\lra{y_{i^*}, t} \\
&\quad\quad + \c\lra{\P\lra{x_{i^*-1}, x_{i^*}}} + \c\lra{\P\lra{y_{i^*}, t}} \\
&\leq \c\lra{\P_{\zeta'_{\pi^*}}} - \c\lra{y_{i^*}, t} + \c\lra{\P\lra{x_{i^*-1}, x_{i^*}}} + \c\lra{\P\lra{y_{i^*}, t}} \\
&\leq \c\lra{\pi^*} - \c\lra{\P_{\zeta_{\pi^*}}\lra{y_{i^*}, t}}+ 2\c\lra{\pi^*} \\
&\quad\quad +\lrA{\alpha+\varepsilon} \c\lra{\P_{\zeta_{\pi^*}}\lra{y_{i^*}, t}} + \frac{4}{L-4}\c\lra{\pi^*} \\
&=\lra{3 + \frac{4}{L-4}}\c\lra{\pi^*} +\lrA{\alpha+\varepsilon - 1} \c\lra{\P_{\zeta_{\pi^*}}\lra{y_{i^*}, t}} \\
&\leq\lra{\alpha + 2 + \frac{4}{L-4} + \varepsilon}\c\lra{\pi^*},
\end{split}
\end{equation}
where the second inequality follows from \eqref{lm4eq1}, \eqref{lm4eq2}, and \eqref{lm4eq4}, and the last inequality from $\alpha+\varepsilon \geq 1$ and $\c\lra{\P_{\zeta_{\pi^*}}\lra{y_{i^*}, t}} \leq \c\lra{\pi^*}$ by the triangle inequality.

The running time is dominated by enumerating $\zeta'_{\pi^*}$ whose length is bounded by $O\lrA{L}$. Furthermore, computing $\P\lra{x_{i^*-1}, x_{i^*}}$ via Lemma~\ref{path} and the $\lrA{\alpha+\varepsilon}$-approximate path $\P'\lra{y_{i^*}, t}$ both run in polynomial time. Therefore, the overall algorithm runs in $N^{O\lrA{L}}$ time. This completes the proof.
\end{IEEEproof}

To formally establish the overall approximation ratio and running time, we unify the results from the above cases. Notice that across all cases, the compressed state sequence $\zeta'_{\pi^*}$ is formed by modifying the tail end of $\zeta_1$ and $\zeta_2$. We can bound the overall length of $\zeta'_{\pi^*}$ rigorously.

\begin{lemma}\label{lem:final_length}
In all cases, the fully compressed state sequence $\zeta'_{\pi^*}$ has length at most $L+3$. 
\end{lemma}
\begin{IEEEproof}
By definition, $\zeta_1$ has exactly $L$ states. In Case 1 and Case 2, $\zeta'_{\pi^*}$ is obtained by compressing the states starting from the end of $\zeta_1$ and extending into $\zeta_2$. The uncompressed prefix of $\zeta_1$ retains its original states. 
By Lemma~\ref{lem:bound}, any legally compressed irreducible state sequence for a path in a bipartite graph has length at most $f\lrA{2} = 8$. However, since $\zeta'_2$ (or the compressed tail) represents a simple path between at most two subsets, it can be further shown that its compressed length is at most $3$. 
Specifically, the remaining compressed suffix contains at most three alternating states between $\V_{3-\theta}$ and $\V_\theta$. Thus, the total length of $\zeta'_{\pi^*}$ cannot exceed the $L$ states from $\zeta_1$ plus the at most $3$ states from the compressed suffix, yielding a maximum length of $L+3$.
\end{IEEEproof}

Finally, we obtain the main result for the \ac{BTSPP} with $k=2$.

\begin{theorem}
For the \ac{BTSPP} with $k=2$, there exists a polynomial-time $\lrA{\alpha+2+\varepsilon}$-approximation algorithm for any constant $\varepsilon>0$.
\end{theorem}
\begin{IEEEproof}
Given any constant $\varepsilon > 0$, we set $L \geq {8}/{\varepsilon} + 4$ so that $4/\lra{L-4} \le \varepsilon/2$, and execute all subroutine algorithms with error parameter ${\varepsilon}/{2}$.

The algorithm enumerates all compressed state sequences of length at most $L+3$.
By Lemma~\ref{lem:final_length}, one of the enumerated sequences is $\zeta'_{\pi^*}$, the compressed state sequence corresponding to an optimal solution.
For each enumerated sequence, we apply the reconstruction subroutines from Lemmas~\ref{lm1}--\ref{lm4}, and return the minimum-cost Hamiltonian path found.

When $\zeta'_{\pi^*}$ is correctly enumerated, the worst-case approximation ratio is determined by Case~3 in Lemma~\ref{lm4}, which satisfies
$\alpha + 2 + 4/\lra{L-4} + \varepsilon/2 \leq \alpha + 2 + \varepsilon/2 + \varepsilon/2 = \alpha + 2 + \varepsilon$.
Hence, the returned solution achieves an $\lrA{\alpha+2+\varepsilon}$-approximation guarantee.

The running time is dominated by enumerating all state sequences, which requires $N^{O\lrA{L}}$ candidates. 
For each enumerated sequence, the reconstruction procedures run in $N^{O\lrA{1}}$ time by Lemmas~\ref{lm1}--\ref{lm4}. 
Since $L=O\lrA{1/\varepsilon}$, the total running time is $N^{O\lrA{1/\varepsilon}}$, which is polynomial for any fixed $\varepsilon>0$.
\end{IEEEproof}

\section{Conclusion}\label{sec:5}
In this paper, we study approximation algorithms for the \ac{BTSP} and the \ac{BTSPP} and develop a general approximation framework for a fixed number of regions. 
By introducing a state-based representation and a compression framework, we show that optimal solutions admit bounded-size structural representations depending only on the number of regions. 
Based on this property, we design polynomial-time $\lrA{\alpha k+1}$- and $\lrA{2k+1}$-approximation algorithms for the \ac{BTSP} and the \ac{BTSPP}, respectively, for any fixed number $k$ of regions. 
These results provide the first constant-factor approximation framework beyond the bipartite case and establish a systematic approach for biased routing problems with multiple regions.

Furthermore, we investigate the bipartite case of the \ac{BTSPP} and develop an improved polynomial-time $\lrA{\alpha+2+\varepsilon}$-approximation algorithm. 
The algorithm exploits additional structural properties of two-region state sequences and combines refined compression techniques with efficient reconstruction procedures to achieve a stronger approximation guarantee than previous approaches.

An interesting direction for future work is to further improve the approximation ratios, especially for the \ac{BTSPP} with multiple regions, and to investigate whether the proposed state compression framework can be extended to other biased routing and related optimization problems.

\section*{Acknowledgment}
AI tools were used only for language polishing and grammatical refinement. All algorithmic and theoretical results were developed independently by the authors without using AI tools.


\end{document}